\documentclass[preprint, numbers, 3p]{elsarticle}

\usepackage[english]{babel}
\usepackage[utf8x]{inputenc}

\usepackage{graphicx} 	
\usepackage{psfrag}
\usepackage[hang,tight,raggedright]{subfigure} 
\usepackage{color}

\usepackage{amsmath}
\usepackage{amsfonts}
\usepackage{amssymb}
\usepackage[amssymb]{SIunits} 

\usepackage{color}
\usepackage{multirow}			
\usepackage[nolist]{acronym} 	

\newtheorem{rmk}{Remark}[section]
\newcommand{\name}[1]{\textit{#1}}

\begin{document}

\title{Modeling the non-trivial behavior of anisotropic beams: a simple Timoshenko beam with enhanced stress recovery and constitutive relations}

\author[a]{Giuseppe Balduzzi\corref{cor}}
\ead{Giuseppe.Balduzzi@tuwien.ac.at}

\author[b]{Simone Morganti}
\ead{simone.morganti@unipv.it}

\author[a]{Josef F\"ussl}
\ead{Josef.Fuessl@tuwien.ac.at}

\author[a]{Mehdi Aminbaghai}
\ead{Mehdi.Aminbaghai@tuwien.ac.at}

\author[c]{Alessandro Reali}
\ead{alessandro.reali@unipv.it}

\author[c]{Ferdinando Auricchio}
\ead{ferdinando.auricchio@unipv.it}

\address[a]{
Institute for Mechanics of Materials and Structures (IMWS), 
Vienna University of Technology, Vienna, Austria}

\address[b]{
Department of Electrical, Computer, and Biomedical Engineering, 
University of Pavia, Pavia, Italy}

\address[c]{
Department of Civil Engineering and Architecture (DICAr), 
University of Pavia, Pavia, Italy}

\cortext[cor]{Corresponding author. \emph{Address}: 
Institute for Mechanics of Materials and Structures (IMWS), 
Vienna University of Technology, Karlsplatz 13/202
A-1040 Vienna, Austria
\emph{Email address}: 
\texttt{Giuseppe.Balduzzi@tuwien.ac.at} 
\emph{Phone}:
 0043 (1) 58 80 12 02 28}

\begin{acronym}[XXXXX]
\acro{FE}   {Finite Element}
\acro{IGA}	{Iso Geometric Analysis}
\acro{IGC}	{Iso Geometric Collocation}
\acro{FEM}  {Finite Element Method}
\acro{ODE}	{Ordinary Differential Equation}
\acro{PDE}	{Partial Differential Equation}
\acro{SV}	{Saint Venant}
\acro{EB}	{Euler Bernoulli}
\acro{KL}	{Kichhoff Love}
\acro{RM}	{Reissner Mindlin}
\acro{TPE}	{Total Potential Energy}
\acro{BC}	{Boundary Condition}
\acro{BVP}	{Boundary Value Problem}
\acro{DOF}	{Degre Of Freedom}
\acro{GLT} 	{Glued Laminated Timber}
\acro{FGM} 	{Functionally Graded Material}
\acro{FSDT}	{First-order Shear Deformation Theory}
\acro{RMVT}	{Reissner Mixed Variational Theorem}
\acro{VABS}	{Variational Asymptotic Beam Sectional Analysis}
\acro{SAFE}	{Semi-Analytical Finite Element}
\acro{GBT}	{Generalized Beam Thory}
\end{acronym}

\begin{abstract}
This paper analyzes the non-trivial influence of the material anisotropy on the structural behavior of an anisotropic multilayer planar beam.
Indeed, analytical results available in literature 
are limited to homogeneous beams and several aspects has not been addressed yet, impeding an in-depth understanding of the mechanical response of anisotropic structural elements.
This paper proposes an effective recovery of stress distribution and an energetically consistent evaluation of constitutive relations to be used within a planar Timoshenko beam model.
The resulting structural-analysis tool highlights the following peculiarities of anisotropic beams:
(i) the axial stress explicitly depends on transversal internal force,  which can weigh up to 30\% on the maximal magnitude of axial stress, and
(ii) the anisotropy influences the beam displacements more than standard shear deformation and even for extremely slender beams.
A rigorous comparison with analytical and accurate 2D Finite Element solutions confirms the accuracy of the proposed approach that leads to errors exceptionally greater than 5\%.
\end{abstract}

\begin{keyword}
Anisotropic multilayer beam 
\sep 
First Order Shear Deformation Theory
\sep 
Analytical solution
\sep 
Beam constitutive relation
\end{keyword}

\maketitle

\section{Introduction} 
\label{s_intro}

Effective analysis and design of timber and composite structures unavoidably require beam and plate models capable to handle the material anisotropy.
Nowadays the need of accurate analysis tools is even more urgent due to the fast development of novel technologies.
As an example, additive manufacturing allows to create structural elements with variable orientation of fibers \citep{gw_14}.
Likewise, laser scanners  detect grain orientation on the surfaces of timber boards, allowing for accurate analyses of glued laminated timber beams \citep{kfse_15} and advanced optimization of structural elements \citep{pklf_18}.

Engineering research has spent great effort in terms of beam and plate modeling within the last decades.
Nevertheless, most of models were derived under the hypothesis of isotropic or, at most, orthotropic material \citep{vgp_12, cc_05}.
As a consequence, several features of anisotropic structural elements are not yet well addressed, in particular when principal directions of the material are not aligned with the beam axis or the plate reference surface.

Limiting the discussion to planar problems, the generalized 2D Hook's law of an isotropic material is represented by a block diagonal matrix. 
Consistently, beam constitutive relations are represented by a diagonal matrix i.e., axial deformation, curvature, and shear deformation uniquely depend on axial internal force, bending moment, and transversal internal force, respectively.
Conversely, anisotropy leads the generalized 2D Hook's law to be represented by a full matrix \citep{l_68}.
As a consequence, also constitutive relations for anisotropic structural elements may be represented by a full matrix i.e., all generalized deformations may depend on all internal forces.

Despite its importance, the above-mentioned problematic was only partially addressed in literature.
\citet{mry_96} proposed a \ac{FSDT} for a planar homogeneous anisotropic beam where a coupling term (mentioned also as coefficient of mutual influence \cite{mvps_10, mv_12}) relates axial deformation with shear force (and vice-versa). 
In the successive years, several researchers \citep{my_96, jnc_02, ql_02, vmp_03, mvps_10} used different approaches for the estimation of the coupling term, reaching slightly different solutions.
As will be discussed in the following, the introduction of a single coupling term allows to define models that are effective only for an extremely limited set of cross-section geometries, while simple and effective models capable to handle more general cases has not been proposed jet.

More recently, \citet{kh_16} have proposed an accurate analysis of the structural behavior of an homogeneous anisotropic planar beam.
The analytical expression for the stress distribution is calculated using the Airy's stress function, analytical expression for deformations is computed using 2D constitutive relations, and the 2D displacement field is recovered using the compatibility \acp{PDE}.
Simple calculations allow to reformulate the obtained analytical expression for 2D displacements and stresses in terms of 1D functions coinciding with internal forces and \ac{FSDT} kinematic parameters.
Taking advantage of this simplification, the authors provide the analytical expression of the \ac{FE} stiffness matrix of the beam.
Analogous stress distributions was obtained also by \citet{h_67}, nevertheless the simplification proposed by \citet{kh_16} highlights that the axial stress explicitly depends on transversal internal force due to the non-trivial constitutive relations of the material.
The above discussed analytical results represent a milestone for the development of effective anisotropic beams.
Nevertheless, the derivation procedure can not be easily generalized to multilayer structures, resulting therefore of limited interest for practitioners. 

An other significant aspect that has to be carefully handled is the length of zones where boundary effects extinguish according to the Saint-Venant principle.
While for a isotropic beam boundary effects are negligible at a distance greater than the maximal size of the cross-section, for an anisotropic beams such a distance depends on the ratio between axial and shear modulus and may be greater than six or seven times the maximal cross-section dimension \cite{ch_77, b_85, hc_18}.
On the one hand, this reduces the effectiveness of beam models and, on the other hand, it introduces further phenomena to be considered in the analysis of structural elements, impeding a straightforward interpretation of both numerical and experimental results.

Nowadays, effective and accurate cross-section analysis tools (e.g., \citep{gw_16a}, Variational Asymptotic Beam Sectional Analysis \citep{yhvc_02, yvhh_02, phd_g_14, gssh_18},  Semi-Analytical Finite Element \citep{dkl_01a,dkl_01b,dkl_01c,dat_10}, Generalized Beam Thory \citep{sc_02}) that may accurately handle the so far introduced problems are available.
Nevertheless, all the cross-section analysis tools are based on auxiliary \acp{PDE} and functionals, impeding an immediate physical understanding of the analysis results.
As a consequence, engineers use the above-mentioned analysis tools as black-boxes.
Furthermore, the scarce awareness about the effects of anisotropy on the structural behavior leads engineers to erroneously believe that coarse adaptations of isotropic beam models are effective \citep{bkf_18}.

This paper proposes a simple planar beam model that effectively describes the linear elastic behavior of anisotropic multi-layer structural elements accounting for the previously introduced issues.
Specifically, the beam model will assume that the beam cross-section behaves rigidly, in analogy to the Timoshenko beam.
On the one hand, this choice limits the accuracy and the applicability of the proposed beam model.
On the other hand, it leads to \acp{ODE} for which an analytical solution can be computed and easily interpreted by simple physical considerations, allowing for a deep understanding of the structural behavior of anisotropic beams.

The main novelty of the developed model is an enhanced and effective stress recovery based on a two-steps iterative procedure.
The former step uses the first 2D constitutive relation for the recovery of axial stress and allows to handle the effects of the anisotropy on the stress distribution.
The latter step uses the horizontal equilibrium \ac{PDE} for the recovery of shear stress, in analogy with standard Jourawsky approach \citep{j_56,b_03}.
Such a procedure allows to identify the non-trivial and explicit dependence of axial stress on transversal internal force and to manage also multi-layer anisotropic beams.
Beam constitutive relations are derived from the stress potential and the outcomes of stress recovery procedure.
Such an approach properly embeds anisotropy effects within the beam model and the effectiveness of the proposed constitutive relation derivation path was already demonstrated for non-prismatic and functionally graded material beams \citep{basfea_16, baaf_17, baf_17, bsaf_17}.

Numerical results will demonstrate that the proposed beam model describes the behavior of anisotropic structural element with a good accuracy, leading to an extremely convenient cost-benefit ratio.
In particular, the proposed beam model effectively predicts: 
(i) the highly non-linear distribution of axial stresses, obtained despite deformations have a linear distribution and the material is linear-elastic,
(ii) the explicit dependency of horizontal stress on transversal internal force and load,
(iii) the fact that the anisotropy influences the beam displacements more than shear deformation.

The outline of the paper is as follows:
Section \ref{s_beam_model} defines the problem and illustrates the beam model \acp{ODE},
Section \ref{s_anal_sol} derives the \acp{ODE} analytical solution,
Sections \ref{s_anal_res} and \ref{s_num_res} discuss some meaningful examples, and
Section \ref{s_conclusion} resumes main properties, advantages, and limitations of the proposed method and delineates future research.

\section{Beam model}
\label{s_beam_model}

This section introduces the 2D problem as well as notations (Section \ref{s_mech_prop}), it discusses the beam compatibility and equilibrium \acp{ODE} (Section \ref{s_equil+comp}), the recovery of cross-section stress distribution (Section \ref{s_stress_rec}), and the beam constitutive relations (Section \ref{s_beam_const_rel}).

\subsection{ 2D problem definition}
\label{s_mech_prop}

\begin{figure}[htbp]
\centering 
\psfrag{O}{\footnotesize $O$}
\psfrag{x}{\footnotesize $x$}
\psfrag{y}{\footnotesize $y$}
\psfrag{p}{\footnotesize $p \left(x\right)$}
\psfrag{q}{\footnotesize $q$}
\psfrag{m}{\footnotesize $m \left(x\right)$}
\psfrag{h}{\footnotesize $h$}
\psfrag{Bh}{\footnotesize $\beta h$}
\psfrag{l}{\footnotesize $l$}
\psfrag{L}{\footnotesize $L$}
\psfrag{omega}{\footnotesize $\Omega$}
\psfrag{theta}{\footnotesize $\theta$}
\includegraphics[width=0.5\columnwidth]{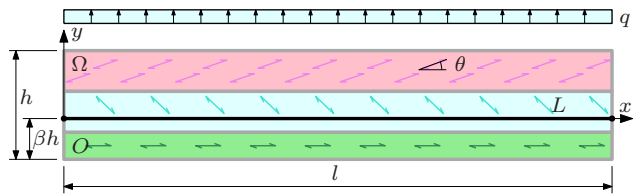}
\caption{\footnotesize{
Anisotropic multilayer beam with arbitrary orientation of principal directions. 
Geometry, coordinate system, dimensions and adopted notations.}} 
\label{f_trave}
\end{figure}

The beam \name{longitudinal axis} $L$ 
and the beam \name{cross-section} $H$ are
closed and bounded subsets of  $x$- and $y$- axes defined as
\begin{equation} 
\label{axis_def}
L := \left\{x \in \left[ 0 , l \right]\right\};
\quad
H := \left\{y \in \left[ - \beta h , \left(1-\beta\right) h \right]\right\}
\end{equation}
where $l$, $h$, and $0<\beta<1$ are the \name{beam length}, the \name{beam thickness}, and a dimensionless parameter defining the distance between the $x$-axis and the lower boundary of the cross-section, respectively.
The \name{beam depth} $b$ denotes the cross-section size along the $z$ coordinate and, in the following, we assume that $b = 1$.
As illustrated in Figure~\ref{f_trave}, the 2D \name{beam body} $\Omega$ is defined as 
\begin{equation} 
\label{beam_body_def}
\Omega := L \times H
\end{equation}
Finally, we assume that the body is slender (i.e., $l \gg h$) and behaves under the hypothesis of plane stress and small displacements.

We introduce
the \name{displacement} vector field $\pmb{s} \left(x,y\right) = \left[ s_x \left(x,y\right) , s_y \left(x,y\right) \right]$,
the \name{stress} tensor field $\pmb{\sigma} \left(x,y\right) = \left[\sigma_{x} \left(x,y\right) , \sigma_{y} \left(x,y\right) , \tau \left(x,y\right)\right]^T$, and
the \name{strain} tensor field $\pmb{\varepsilon} \left(x,y\right) = \left[\epsilon_{x}  \left(x,y\right) , \epsilon_{y} \left(x,y\right) , \gamma_{xy} \left(x,y\right)\right]^T$ using the engineering notation.
Furthermore,  a \name{distributed load} $\pmb{f} = \left[ f_x  , f_y \right]$ is applied within the domain and suitable \acp{BC} are assigned on the boundary of domain $\Omega$.
The 2D compatibility \acp{PDE} read
\begin{subequations}\label{compat_eq}
\begin{align}
\epsilon_{x} \left(x,y\right) = & s_{x,x} \left(x,y\right)
\label{comp_epsx}\\
\epsilon_{y} \left(x,y\right) = & s_{y,y} \left(x,y\right)
\label{comp_epsy}\\
\gamma_{xy} \left(x,y\right) = & \frac{1}{2} \left( s_{x,y} \left(x,y\right) + s_{y,x} \left(x,y\right) \right)
\label{comp_gamm}
\end{align}
\end{subequations}%
where the notation $\left(\cdot\right),i$ for $i=x,y$ represents partial derivatives.
The 2D equilibrium \acp{PDE} read
\begin{subequations}\label{equil_eq}
\begin{align}
\sigma_{x,x} \left(x,y\right) + \tau_{,y} \left(x,y\right) = & -f_x \left(x,y\right)
\label{equil_x}\\
\tau_{,x} \left(x,y\right) + \sigma_{y,y} \left(x,y\right) = & -f_y \left(x,y\right)
\label{equil_y}
\end{align}
\end{subequations}

The beam is made of a linear-elastic and anisotropic material.
As represented in Figure \ref{f_trave}, the material properties do not depend on the beam axis $x$ coordinate and are piecewise constant within the beam thickness.
Following the notation introduced by \citet{mry_96}, the 2D anisotropic constitutive relations can be represented as

\begin{subequations}\label{mat_const_rel}
\begin{align}
\epsilon_x \left(x,y\right) = & \frac{\sigma_{x}\left(x,y\right)}{E_{xx}\left(y\right)} + \frac{\sigma_{y}\left(x,y\right)}{E_{xy}\left(y\right)} + \frac{\tau\left(x,y\right)}{G_x\left(y\right)}
\label{mat_const_rel_x}
\\
\epsilon_y \left(x,y\right) = & \frac{\sigma_{x}\left(x,y\right)}{E_{xy}\left(y\right)} + \frac{\sigma_{y}\left(x,y\right)}{E_{yy}\left(y\right)} + \frac{\tau\left(x,y\right)}{G_y\left(y\right)}
\label{mat_const_rel_y}
\\
\gamma_{xy} \left(x,y\right) = & \frac{\sigma_{x}\left(x,y\right)}{G_{x}\left(y\right)} +  \frac{\sigma_{y}\left(x,y\right)}{G_{y}\left(y\right)} +  \frac{\tau\left(x,y\right)}{G\left(y\right)}
\label{mat_const_rel_xy}
\end{align}
\end{subequations}
The coefficients of the material constitutive relations can be collected in a
matrix $\pmb{D}$ that is defined as
\begin{equation} 
\label{const_rel_rotation}
\pmb{D}\left(y\right) =
\left[
\begin{array}{ccc}
\frac{1}{E_{xx}\left(y\right)} & \frac{1}{E_{xy}\left(y\right)} & \frac{1}{G_x\left(y\right)} \\
\frac{1}{E_{xy}\left(y\right)} & \frac{1}{E_{yy}\left(y\right)} & \frac{1}{G_y\left(y\right)} \\
\frac{1}{G_{x}\left(y\right)}  & \frac{1}{G_{y}\left(y\right)}  & \frac{1}{G\left(y\right)}
\end{array}
\right] =
\pmb{R}^T\left(y\right)
\left[
\begin{array}{ccc}
\frac{1}{{E}_{11}\left(y\right)} & -\frac{\nu\left(y\right)}{{E}_{11}\left(y\right)} & 0 \\
-\frac{\nu\left(y\right)}{{E}_{11}\left(y\right)} & \frac{1}{E_{22}\left(y\right)} & 0 \\
0 & 0 & \frac{1}{G_{12}\left(y\right)}
\end{array}
\right]
 \pmb{R}\left(y\right)
\end{equation}
where
$\pmb{R}\left(y\right)$ reads
\begin{equation}
\pmb{R}\left(y\right) = 
\left[
\begin{array}{ccc}
\cos ^{2} \left( \theta\left(y\right) \right) & 
\sin ^{2} \left( \theta\left(y\right) \right) & 
2 \sin \left( \theta\left(y\right) \right) \cos \left( \theta\left(y\right) \right) 
\\ 
\sin ^{2} \left( \theta\left(y\right) \right) & 
\cos ^{2} \left( \theta\left(y\right) \right) & 
-2 \sin \left( \theta\left(y\right) \right) \cos \left( \theta\left(y\right) \right) 
\\ 
-\sin \left( \theta\left(y\right) \right) \cos \left( \theta\left(y\right) \right) &
\sin \left( \theta\left(y\right) \right) \cos \left( \theta\left(y\right)  \right) & 
\cos ^{2} \left( \theta\left(y\right) \right) - \sin ^{2} \left( \theta\left(y\right) \right)
\end{array}
\right]
\end{equation}
${E}_{11}\left(y\right) , E_{22}\left(y\right) , G_{12}\left(y\right)$, and $\nu\left(y\right)$ are the parameters defining the mechanical properties of the material with respect to the principal directions.
The quantity $\theta\left(y\right)$, with $-\pi/2 < \theta\left(y\right) < \pi/2$, is the rotation of the principal direction of the material with respect to the $x$-axis.

{\rmk 
\label{r_D_even_odd} 
Due to the definition \eqref{const_rel_rotation}, $E_{xx} \left(y\right)$, $E_{xy} \left(y\right)$, $E_{yy} \left(y\right)$, and $G \left(y\right)$ are even functions of the material principal direction rotation $\theta\left(y\right)$ whereas the material coupling terms $G_{x} \left(y\right)$ and $G_{y} \left(y\right)$ are odd. }

\subsection{Compatibility and equilibrium \acp{ODE}}
\label{s_equil+comp}

For convenience, we define the \name{axial stiffness} $A^*$, 
the dimensionless parameter $\beta$ introduced in Equation \eqref{axis_def}, and the \name{bending stiffness} $I^*$
\begin{equation} 
\label{area_centlin_inertia}
A^*  = \int_{H}  E_{xx} \left(y\right) dy;
\quad
\beta  = \cfrac{1}{h A^* }\int_{H}  E_{xx}\left(y \right) y dy;
\quad
I^*  =  \int_{H} E_{xx}\left(y\right)  y^2 dy
\end{equation}

{\rmk 
\label{r_centerline} 
Due to Definition \eqref{area_centlin_inertia}, the origin of the adopted Cartesian coordinate system  $O$ coincides with the so-called stiffness centroid that is equal to the cross-sectional geometric centroid only if the cross-section is symmetric and, in particular, when the beam is homogeneous, as discussed also by \citet{dkl_01b}.}

As usual for standard Timoshenko beam models, the 2D displacement field $\pmb{s}\left(x,y\right) = \left[s_x\left(x,y\right), s_y\left(x,y\right)\right]^T$ is represented in terms of three 1D functions, indicated as \name{axial displacement} $u\left(x\right)$, \name{cross-section rotation} $\phi\left(x\right)$, and \name{transversal displacement} $v\left(x\right)$.
Therefore, the displacement field components are approximated as follows
\begin{subequations} 
\label{kin_ass}
\begin{align}
s_x\left(x,y\right) \approx & 
u \left(x\right) - y \phi \left(x\right)\\
s_y\left(x,y\right) \approx & 
v\left(x\right)
\end{align}
\end{subequations}

Introducing the \name{generalized strains} defined as the \name{axial strain} $\epsilon \left(x\right)$, the \name{curvature} $\chi \left(x\right)$, and the \name{shear strain} $\gamma \left(x\right)$,
the beam compatibility is expressed through the following \acp{ODE}
\begin{subequations} 
\label{disp_rec}
\begin{align}
\epsilon \left(x\right) & = u '\left(x\right) \label{u_rec}\\
\chi \left(x\right)  &= \phi'\left(x\right) \label{fi_rec} \\
\gamma \left(x\right) & = v'\left(x\right) - \phi\left(x\right) \label{v_rec}
\end{align}
\end{subequations}
where the notation $\left(\cdot\right)'$ denotes derivatives with respect to $x$.
{\rmk 
\label{r_approx} 
In light of Remark \ref{r_centerline}, kinematic approximation \eqref{kin_ass} differs from standard Timoshenko one.
In fact, $u \left(x\right)$ represents the axial displacement of the stiffness centroids and, in general,  it does not coincide with the mean value of the cross-section axial displacements (i.e., $u \left(x\right) = s_x \left(x,0\right) \neq 1/h \int_{H} s_x \left(x,y\right) dy$).
Similarly, $\epsilon \left(x\right)$ is not the mean value of the axial strain evaluated within the cross-section, but just represents the axial elongation evaluated at $y=0$ (i.e., $\epsilon \left(x\right) \neq 1/h \int_{H} \partial s_{x} \left(x,y\right)/\partial x \, dy$).}

We introduce the \name{axial internal force} $N \left(x\right)$, the \name{bending moment} $M \left(x\right)$, and the \name{transversal internal force} $V \left(x\right)$ defined as
\begin{gather} 
\label{stress_resultant}
N \left( x \right) = \int_H \sigma_{x} \left(x,y\right) dy;
\quad
M \left( x \right) = \int_H \sigma_{x} \left(x,y\right) \left(-y\right) dy;
\quad
V \left( x \right) = \int_H \tau \left(x,y\right) dy
\end{gather}
Furthermore, we assume that $f_x =0$ and we introduce the transversal load $q$ %
\begin{equation}
q = \int_H f_y  dy = hf_y
\end{equation}
Considering the axial, rotational, and transversal equilibrium of a infinitesimally long beam-segment, the equilibrium \acp{ODE} read
\begin{subequations} 
\label{beam_equil}
\begin{align}
N'\left(x\right) & = 0 
\label{beam_equil_h}
\\
M'\left(x\right) & = - V\left(x\right)  
\label{beam_equil_m}
\\
V'\left(x\right) & = - q
\label{beam_equil_v}
\end{align}
\end{subequations}

\subsection{Stress recovery}
\label{s_stress_rec}

The stress recovery is based on a recursive procedure that leads to define a distribution of stresses that satisfies the first constitutive relation \eqref{mat_const_rel_x} and the first equilibrium \ac{PDE} \eqref{equil_x}.
Conversely, we assume that transversal stress vanishes i.e., $\sigma_{y} \left(x,y\right) =0$, aiming at the maximal simplicity of the model.

In order to set the recursive procedure up, it is convenient to isolate Equations \eqref{mat_const_rel_x} and  \eqref{equil_x} for the variables $\sigma_{x} \left(x,y\right)$ and $\tau \left(x,y\right)$, respectively
\begin{equation}
\label{const_rel_sx}
\sigma_{x} \left(x,y\right) = 
E_{xx} \left(y\right) \epsilon_{x} \left(x,y\right) - 
\frac{E_{xx} \left(y\right)}{G_{x} \left(y\right)}\tau \left(x,y\right)
\end{equation}
\begin{equation} 
\label{shear_p_rec}
\tau\left(x,y\right) =  -\int_{-\beta h}^{y} \sigma_{x,x} \left(x,\hat{y}\right) d\hat{y} 
\end{equation}

\begin{figure}[htbp]
\centering 
\includegraphics[width=0.5\columnwidth]{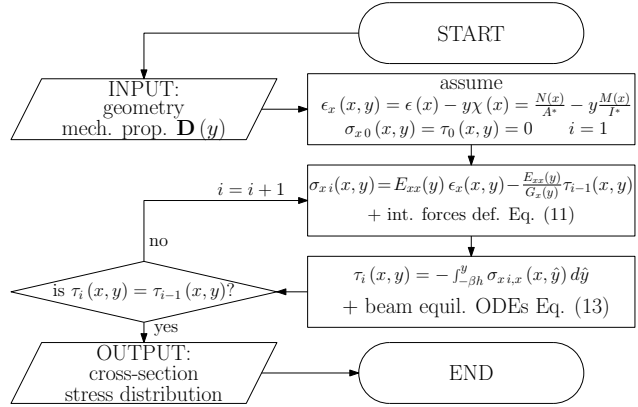}
\caption{\footnotesize{
Flow chart of the iterative procedure adopted for the stress recovery.}} 
\label{f_stress_recovery}
\end{figure}

The iterative procedure is resumed in Figure \ref{f_stress_recovery} and leads to the following distribution of stresses
\begin{subequations}
\label{stress_def}
\begin{align}
\label{sigx_def}
\sigma_{x} \left(x,y\right) = 
d_{\sigma_x}^N \left(y\right) N \left(x\right) + 
d_{\sigma_x}^M \left(y\right) M \left(x\right) + 
d_{\sigma_x}^V \left(y\right) V \left(x\right) + 
d_{\sigma_x}^q \left(y\right) q 
\\
%
\label{tau_def}
\tau \left(x,y\right) = 
d_{\tau}^V \left(y\right) V \left(x\right) + 
d_{\tau}^q \left(y\right) q
\end{align}
\end{subequations}
where
\begin{subequations}
\begin{align}
d_{\sigma_x}^N\left(y\right) & = \frac{E_{xx}\left(y\right)}{A^*}
\label{b_sx_N}
\\ 
d_{\sigma_x}^M\left(y\right) & = -\frac{E_{xx}\left(y\right)}{I^*} y 
\label{b_sx_M}
\\
d_{\sigma_x}^V\left(y\right) & = 
-
\frac{E_{xx} \left(y\right)}{G_{x} \left(y\right)} d_{\tau}^V\left(y\right) +
\int_H \frac{E_{xx} \left(y\right)}{G_{x} \left(y\right)} d_{\tau}^V\left(y\right) dy \, d_{\sigma_x}^N\left(y\right)  
- \int_H \frac{E_{xx} \left(y\right)}{G_{x} \left(y\right)} d_{\tau}^V\left(y\right) y dy \, d_{\sigma_x}^M\left(y\right)
\label{b_sx_V}
\\
d_{\sigma_x}^{q}\left(y\right) & = 
- \frac{E_{xx} \left(y\right)}{G_{x} \left(y\right)} d_{\tau}^{q}\left(y\right) +
\int_H \frac{E_{xx} \left(y\right)}{G_{x} \left(y\right)} d_{\tau}^{q}\left(y\right) dy \, d_{\sigma_x}^N\left(y\right)  - 
\int_H \frac{E_{xx} \left(y\right)}{G_{x} \left(y\right)} d_{\tau}^{q}\left(y\right) y dy \, d_{\sigma_x}^M\left(y\right)
\label{b_sx_q}
\\ 
d_{\tau}^V\left(y\right) & =  \int_{-\beta h}^{y} d_{\sigma}^M \left( \hat{y} \right)  d\hat{y}
\label{b_tau_V}
\\ 
d_{\tau}^q\left(y\right) & = \int_{-\beta h}^{y} d_{\sigma}^V \left( \hat{y} \right)  d\hat{y}
\label{b_tau_q}
\end{align}
\end{subequations}
{\rmk 
\label{r_stress_rec} 
In Definitions \eqref{b_sx_V} and \eqref{b_sx_q}, the second and the third addends satisfy Equation \eqref{stress_resultant}, maintaining standard physical meaning of beam model variables.
}

Equations \eqref{sigx_def} highlights that axial stress $\sigma_{x}$ explicitly depends on transversal internal force $V \left(x\right)$ and load $q$.
Similarly, also the shear stress $\tau$ explicitly depends on the transversal load $q$.
To the authors' knowledge, only \citet{kh_16} and \citet{h_67} obtained similar dependencies, but their analysis was limited to homogeneous beams.
Furthermore, $d_{\sigma_x}^V$ and $d_{\sigma_x}^q$ depend on $E_{xx} / G_{x}$ and $E_{xx}^2 / G_{x}^2$, respectively (see Equation \eqref{b_sx_V} and \eqref{b_sx_q}), and similar coefficients was reported also by \citet{kh_16}.

On the one hand, similarity of Equations \eqref{sigx_def} and \eqref{tau_def} with analytical solutions reported in \citep{kh_16} and \citep{h_67} indicates that the procedure summarized in Figure \ref{f_stress_recovery} may be effective.
On the other hand, non-trivial dependency of $\sigma_{x}$ on transversal internal force $V \left(x\right)$ indicates that stress recovery procedures developed for isotropic or orthotropic structural elements available in literature \citep{dasar_17, tfbr_17} and implemented in most of structural analysis commercial softwares can lead to coarse results.

\subsection{Beam constitutive relations}
\label{s_beam_const_rel}

To complete the Timoshenko-like beam model, simplified constitutive relations have to be defined.
To this aim, we introduce the stress potential
\begin{equation} 
\label{stress_potential}
\Psi^*\left(x,y\right) = \frac{1}{2} \,
\pmb{\sigma}^T\left(x,y\right) \cdot \pmb{D}\left(y\right) \cdot \pmb{\sigma}\left(x,y\right) = \frac{1}{2} \left( \frac{\sigma_{x}^2\left(x,y\right) }{E_{xx}\left(y\right) } +  \frac{\tau^2\left(x,y\right)}{G\left(y\right)} + 2 \frac{\sigma_{x}\left(x,y\right)\tau\left(x,y\right) }{G_{x}\left(y\right) }\right)
\end{equation}
Substituting the stress recovery relations \eqref{stress_def} into Equation \eqref{stress_potential}, the generalized strains result as the cross-section integral of the derivatives of the stress potential with respect to the corresponding internal forces, reading
\begin{subequations}  
\label{const_rel}
\begin{align}
\epsilon \left(x\right) 
= &
\int_{H}
\frac{\partial \Psi^*\left(x,y\right)}{\partial N\left(x\right)}
 dy =
\epsilon_N N \left(x\right) + \epsilon_M M \left(x\right) + \epsilon_V V \left(x\right)
+
\epsilon^q q 
\\
\chi \left(x\right) 
= &
\int_{H}
\frac{\partial \Psi^*\left(x,y\right)}{\partial M\left(x\right)}
 dy =
\chi_N N \left(x\right) + \chi_M M \left(x\right) + \chi_V V \left(x\right)
+
\chi^q q 
\\
\gamma \left(x\right) 
= &
\int_{H}
\frac{\partial \Psi^*\left(x,y\right)}{\partial V\left(x\right)}
 dy =
\gamma_N N \left(x\right) + \gamma_M M \left(x\right) + \gamma_V V \left(x\right)
+
\gamma^q q 
\end{align}
\end{subequations}
with
\begin{subequations}
\begin{align}
\epsilon_N & = \int_{H} 
\frac{\left( d_{\sigma_x}^N\left(y\right) \right)^2}{E_{xx}\left(y\right)}  dy
\label{eps_N}
\\
\epsilon_M = \chi_N & =  \int_{H} 
\frac{d_{\sigma_x}^N\left(y\right)  d_{\sigma_x}^M\left(y\right) }{E_{xx}\left(y\right)}  dy = 0 
\label{eps_M}
\\
\epsilon_V = \gamma_N & =
\int_{H}
\frac{d_{\sigma_x}^N\left(y\right) d_{\sigma_x}^V\left(y\right)}{E_{xx}\left(y\right)} dy +  
\int_{H}
\frac{d_{\sigma_x}^N\left(y\right) d_{\tau}^V\left(y\right)}{G_{x}\left(y\right)} dy 
\label{eps_V}
\\
\chi_M  & =  \int_{H}  \frac{\left( d_{\sigma_x}^M\left(y\right) \right)^2}{E_{xx}\left(y\right)}  dy
\label{chi_M}
\\
\chi_V = \gamma_M & =  
\int_{H}  \frac{d_{\sigma_x}^M\left(y\right) d_{\sigma_x}^V\left(y\right)}{E_{xx}\left(y\right)} dy +
\int_{H}  \frac{d_{\sigma_x}^M\left(y\right) d_{\tau}^V\left(y\right)}{G_{x}\left(y\right)} dy
\label{chi_V}
\\
\gamma_V & = 
\int_{H} \frac{\left( d_{\sigma_x}^V\left(y\right) \right)^2}{E_{xx}\left(y\right)} dy +
\int_{H} \frac{d_{\sigma_x}^V d_{\tau}^V\left(y\right)}{G_{x}\left(y\right)} dy +
\int_{H} \frac{\left( d_{\tau}^V\left(y\right) \right)^2}{G\left(y\right)} dy
\label{gam_V}
\\
\epsilon^q & = 
\int_{H} \frac{d_{\sigma_x}^N\left(y\right) d_{\sigma_x}^q \left(y\right) }{E_{xx}\left(y\right)} dy +
\int_{H} \frac{d_{\sigma_x}^N\left(y\right) d_{\tau}^q\left(y\right)}{G_{x}\left(y\right)} dy
\label{eps_p}
\\
\chi^q & =
\int_{H} \frac{d_{\sigma_x}^M\left(y\right) d_{\sigma_x}^q \left(y\right) }{E_{xx}\left(y\right)} dy +
\int_{H} \frac{d_{\sigma_x}^M\left(y\right) d_{\tau}^q\left(y\right)}{G_{x}\left(y\right)} dy
\label{chi_p}
\\
\gamma^q & =  
\int_{H} \frac{d_{\sigma_x}^V\left(y\right) d_{\sigma_x}^q \left(y\right) }{E_{xx}\left(y\right)} dy +
\int_{H} \frac{d_{\sigma_x}^V\left(y\right) d_{\tau}^q\left(y\right)}{G_{x}\left(y\right)} dy +
\int_{H} \frac{d_{\sigma_x}^q\left(y\right) d_{\tau}^V\left(y\right)}{G_{x}\left(y\right)} dy +
\int_{H} \frac{d_{\tau}^V \left(y\right) d_{\tau}^V\left(y\right)}{G\left(y\right)} dy
\label{gam_p}
\end{align}
\end{subequations}

Introducing the definitions of stress distributions \eqref{sigx_def} into Definition \eqref{eps_M}, we obtain that $\epsilon_M = \chi_N = 0$ due to the choice of the origin of the Cartesian coordinate system introduced in Equation \eqref{area_centlin_inertia}.
Conversely, the transversal internal force $V \left(x\right)$ produces not only shear deformation $\gamma \left(x\right)$, but also axial strain $\epsilon \left(x\right)$ and curvature $\chi \left(x\right)$ since $\epsilon_V \neq 0$ and $\chi_V \neq 0$.
Equations \eqref{eps_N} and \eqref{chi_M} lead to a definition of axial and bending stiffness analogous to the one obtained for isotropic beams.
Conversely, Equation \eqref{gam_V} highlights that the shear stiffness of anisotropic beams $\gamma_V$ depends not only on shear modulus $G \left(y\right)$, but also on both the axial modulus of elasticity $E_{xx} \left(y\right)$ and the coupling term $G_{x} \left(y\right)$. 
Finally, all the deformations explicitly depend on the transversal load $q$ (see Equation \eqref{const_rel}).

Such an deep influence of the material anisotropy on beam constitutive relations is ignored by most of the literature.
To the authors' knowledge, the coefficients $\epsilon_V =\gamma_N$ was analyzed only in \citep{mry_96, my_96, jnc_02, ql_02, vmp_03}.
Conversely, the existence of the coefficients $\chi_V = \gamma_M$ was mentioned by \citep{dkl_01b} in the framework of the derivation of an enhanced 3D beam model, but their influence on the beam structural response was never analyzed.

{\rmk 
\label{r_bondary_effects}
The extremely simple assumptions on kinematics \eqref{kin_ass} do not allow to tackle any higher order and boundary effects, as usual for all \ac{FSDT}. 
Therefore, the proposed beam model has not the capability to describe deformation of the cross-section and the phenomena that occur in the neighborhood of constraints and concentrated loads.}

\section{\acp{ODE} analytical solution}
\label{s_anal_sol}

\begin{figure}[htbp]
\centering
\psfrag{l}{\footnotesize $l$}
\psfrag{h1}{\footnotesize $h_1$}
\psfrag{h2}{\footnotesize $h_2$}
\psfrag{hh}{\footnotesize $h$}
\psfrag{p}{\footnotesize $q$}
\psfrag{can}{\footnotesize cantilever}
\psfrag{dub}{\footnotesize doubly-clamped}
\includegraphics[width=0.5\columnwidth]{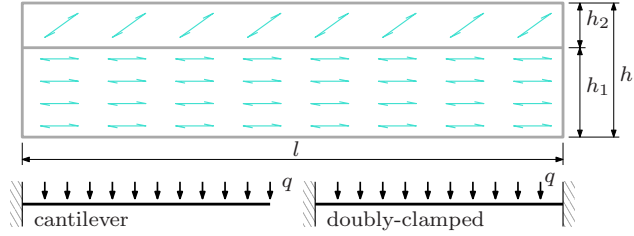}
\caption{\footnotesize{
Bi-layer anisotropic cantilever. 
Geometry, loads, and \acp{BC}.}} 
\label{f_cantilever}
\end{figure}

This section discusses the analytical solution of beam model \acp{ODE} \eqref{equil_eq}, \eqref{disp_rec}, and \eqref{const_rel} for the two-layer cantilever depicted in Figure \ref{f_cantilever}. 
The two layers are made of the same anisotropic material and their thicknesses are  $h_1 = \alpha h$ and $h_2 = \left(1-\alpha\right) h$ with $0 \leq \alpha \leq 1$.
In the bottom layer, material principal direction is aligned with the beam axis, therefore $E_{xx} = E_{11}$, $G = G_{12}$, and $1/G_{x} = 0$.
In the top layer material principal direction is rotated with respect to the beam axis of an angle $\theta$, therefore $E_{xx} = E_{11} / \mu$ and $G = G_{12} / \kappa$. 
The dimensionless parameters $\mu$ and $\kappa$ account for the reduction of axial and shear modulus due to the rotation $\theta$ of the material principal direction \eqref{const_rel_rotation}.
They are defined in Appendix \ref{a_mat_coeff}, together with the material coupling term $G_{x}$.
For $\alpha = 1$, the beam reduces to an homogeneous orthotropic beam with material orientation aligned with beam axis.
Conversely, for $\alpha = 0$, the beam reduces to an homogeneous anisotropic beams, similar to the one analyzed by \citep{mry_96, my_96, jnc_02, ql_02, vmp_03, h_67, kh_16}.
Finally, aiming at the maximal simplicity of the analytical solution, we are going to neglect the influence of transversal load on the beam deformation (i.e., we assume $\epsilon^q = \chi^q = \gamma^q = 0$).


The coefficients of beam constitutive relations introduced in Definition \eqref{const_rel} read
\begin{equation}
\label{const_rel_coeff}
\epsilon_N = 
\frac{P_1}{E_{11} h};
\quad 
\epsilon_V = \gamma_N =
- \frac{P_3}{G_{x} h};
\quad  
\chi_M = 
\frac{12 P_2}{E_{11} h^3};
\quad
\chi_V = \gamma_M =
- \frac{18 P_4}{G_{x} h^2};
\quad
\gamma_V =
\frac{6 P_5}{5 G_{12} h} 
 + 
\frac{P_6 E_{11}}{5 G^2_{x} h}
\end{equation}
where the dimensionless coefficients $P_i$ for $i= 1 \dots 6$ (reported in Appendix \ref{a_p}) account for the influence of geometry and material on the structural element stiffness.

The solution of \acp{ODE} \eqref{disp_rec}, \eqref{equil_eq}, and \eqref{const_rel} leads to the following analytical expressions for beam model variables
\begin{equation}
\label{ode_sol}
\begin{split}
N \left( x \right) = 
&
C_6
\\
V \left( x \right) = 
&
-q x + C_5
\\
M \left( x \right) =
&
\frac{q x^2}{2} - C_5 x + C_3 
\\
\phi \left( x \right) = 
&
\overbrace{\frac{12 P_2}{E_{11} h^3} \left(\frac{q x^3}{6}  -\frac{C_5 {x}^{2}}{2}  + C_3 x \right)}^{\phi_{EB} \left( x \right)}
+
\overbrace{\frac{18 P_4}{G_{x} h^2} \left(-\frac{q x^2}{2}  + C_5 x \right)}^{\phi_{c} \left( x \right)} + C_2
\\
v \left( x \right) = 
& 
\overbrace{\frac{12 P_2}{E_{11} h^3} \left( \frac{q x^4}{24} - \frac{C_5 {x}^{3}}{6} +
\frac{C_3 {x}^{2}}{2} \right)}^{v_{EB} \left( x \right)}
+ \overbrace{\frac{6 P_5}{5 G_{12} h} \left( - \frac{q x^2}{2} + C_5 x \right)}^{v_{T} \left( x \right)}
\\
+ &\overbrace{\frac{P_3}{G_{x} h} C_6 x -\frac{18 P_4}{G_{x} h^2} C_3 x }^{v_{c} \left( x \right)}
+ \overbrace{\frac{P_6 E_{11}}{ G_{x}^2 h} \left( - \frac{q x^2}{2} + C_5 x \right)}^{v_{r} \left( x \right)}
+C_2 x+C_1
\\
u  \left( x \right) = 
&
\overbrace{\frac{P_1}{E_{11} h} C_6 x}^{u_{EB} \left( x \right)}
- \overbrace{\frac{P_3}{G_{x} h}\left(-\frac{q x^2}{2} + C_5 x \right)}^{u_{c} \left( x \right)}
+C_4
\end{split}
\end{equation}
where $C_i$ for $i=1 \dots6$ depend on \acp{BC}.
Notations $\left(\cdot \right)_{EB}$, $\left(\cdot \right)_{T}$, $\left(\cdot \right)_{c}$, and  $\left(\cdot \right)_{r}$ highlight dependency of addends on the axial stiffness $E_{11}$, the shear stiffness $G_{12}$, the coupling term $G_{x}$, and the ratio $G_{x}^2/E_{11}$, respectively.
Furthermore, few calculations allow to conclude that the addends denoted as $\left(\cdot \right)_{EB}$ coincide with the solution of the \ac{EB} beam theory whereas the addend denoted as $\left(\cdot \right)_{T}$ coincides with the shear deformation considered by Timoshenko beam theory.

Considering a cantilever (see Figure \ref{f_cantilever}), the following \acp{BC} have to be enforced
\begin{equation}
\label{bc_cantilever}
u \left( 0 \right) = 0; \quad
\phi \left( 0 \right) = 0; \quad
v \left( 0 \right) = 0; \quad
N \left( l \right) = 0; \quad
M \left( l \right) = 0; \quad
V \left( l \right) = 0
\end{equation}
Requiring \acp{ODE} solution \eqref{ode_sol} to satisfy \acp{BC} \eqref{bc_cantilever} leads to determine the following value of $C_i$ for $i=1 \dots 6$
\begin{equation}
C_1 = C_2 = C_4 = C_6 = 0; \quad 
C_3 = \frac{q l^2}{2}; \quad
C_5 = q l
\end{equation}
Finally, introducing the dimensionless parameter $\lambda = l/h$, the maximal transversal displacement of the beam reads
\begin{equation} 
\label{v_max}
v \left( l \right) = 
v_{EB} \left( l \right) +
v_{T} \left( l \right) +
v_{c} \left( l \right) +
v_{r} \left( l \right)
=
\frac{3 q l \lambda^3}{2 E_{11}} Q_1 
+ \frac{3 q l \lambda}{G_{12}} Q_2 
- \frac{9 q l \lambda^2}{G_{x}} Q_3
+ \frac{ q l \lambda }{G_{x}}\frac{ E_{11}}{G_{x}} Q_4 
\end{equation}
where the dimensionless coefficients $Q_i$ for $i= 1 \dots 4$ are reported in Appendix \ref{a_q_c}.

Equation \eqref{v_max} shows that the maximal transversal displacement $v \left( l \right)$ is the sum of four terms.
The first addend $v_{EB} \left( l \right)$ depends on the Young's modulus along the principal direction $E_{11}$ and, for $\alpha = 1$, it corresponds to the classical \ac{EB} solution.
The second addend $v_{T} \left( l \right)$ depends on the shear modulus $G_{12}$ and, for $\alpha = 1$, it coincides with the contribution due to shear deformation handled by the Timoshenko beam model.
The third term $v_{c} \left( l \right)$ depends on the material coupling term $G_{x}$ and its existence is just a consequence of the fact that material principal directions are not aligned with the beam axis.
The fourth term  $v_{r} \left( l \right)$ depends on the material coupling term $G_{x}$ and on the ratio $E_{11}/G_{x}$ that appears in the definition of axial stress (see Equation \eqref{b_sx_V}).

Looking at Equation \eqref{v_max} from a different perspective, the first addend $v_{EB} \left( l \right)$ depends on $\lambda ^3$, the second $v_{T} \left( l \right)$ and the fourth  $v_{r} \left( l \right)$ ones depend on $\lambda$, and the third one  $v_{c} \left( l \right)$ depends on $\lambda ^2$.
On the one hand, the so far highlighted result is conformal to what stated in standard literature. 
Shear deformation $v_{T} \left( l \right)$ has a negligible influence on the total displacement of the beam for slender beams (i.e., for $\lambda \gg 1$).
On the other hand, the third term $v_{c} \left( l \right)$ can weigh on the total displacement more than the shear deformation whereas the forth term $v_{r} \left( l \right)$ can have an influence similar to the shear deformation.
To the authors' knowledge, the existence of terms  $v_{c} \left( l \right)$ and  $v_{r} \left( l \right)$ was never mentioned in the literature and their role will be analyzed in the following section.

Finally, it is worth mentioning that
\begin{itemize} 
\item
stress distributions \eqref{stress_def} reduce to linear and quadratic functions for $\alpha = 1$ and $1 / G_{x} =0$, as usual in homogeneous prismatic beams,
\item
$\epsilon_N = 1/ \left( E_{11} h \right)$, $\epsilon_V = \gamma_N = \chi_V = \gamma_M = 0$,  $\chi_M =12/ \left( h^3 E_{11} \right)$, and $\gamma_V = 6/ \left(5G_{12}h \right)$ for $\alpha = 1$, analogously to homogeneous prismatic isotropic beams,
\item
 $\epsilon_N = \mu/ \left( E_{11} h \right)$, $\epsilon_V = \gamma_N = 1/\left(G_{x}h \right)$, $\chi_M =12 \mu/\left( h^3 E_{11}\right)$, $\chi_V = \gamma_M = 0$, and $\gamma_V = 6\kappa/\left(5G_{12}h\right)$ for $\alpha = 0$, similarly to anisotropic beam model proposed by \citet{mry_96}.
\end{itemize}
confirming that the presented beam model can recover analytical solutions already available in literature.

\section{Comparison with analytical solution, simply-supported homogeneous beam, }
\label{s_anal_res}

This section compares the solution of the beam model discussed in Section \ref{s_beam_model} with the analytical solution derived in \citep{kh_16} for a simply supported homogeneous beam.
Numerical results are obtained assuming the following parameters 
\begin{equation}
\label{mech_prop_homog_beam}
\begin{split}
& h = 0.2 \, \milli\meter; \quad
l = 2 \, \milli\meter; \quad
q = 1 \, \newton/\milli\meter; \quad
\alpha = 0
\\
E_{11} = 10^4 \, \mega & \pascal; \quad
E_{22} = 5 \cdot10^2 \, \mega\pascal; \quad
G = 10^3 \, \mega\pascal; \quad
\nu = 0.25
\end{split}
\end{equation}
It is worth mentioning that the material is highly anisotropic: $E_{11}/E_{22} = 20$ and $E_{11}/G = 10$.
Such a choice aims at magnifying the effects of both shear deformation and coupling on the behavior of the structural element, allowing for a more accurate discussion of the beam model effectiveness.

Being $\psi \left(x\right)$ a beam model variable, the solution computed by means of \acp{ODE} \eqref{disp_rec}, \eqref{equil_eq}, and \eqref{const_rel} is denoted in the following as $\psi^{mod}$.
Conversely, the reference solution $\psi^{ref}$ is computed using the analytical expressions reported in \citep{kh_16}.

Assuming $\theta = 45 \, \deg$ the rotation of the constitutive relation \eqref{const_rel_rotation} leads to set
\begin{equation}
\begin{aligned}
E_{xx} & = E_{yy} =  1.904 \, \giga\pascal ; \quad & 
E_{xy} & = 40.000 \, \giga\pascal ; \quad &
G_{x}  & = G_{y}   = -1.052 \, \giga\pascal ; \quad & 
G      & =     0.322 \, \giga\pascal 
\end{aligned}
\end{equation}

\begin{figure}[htbp]
\centering
\subfigure[axial stress, $x=1$]{
\label{f_homog_sx1}
\includegraphics[width=0.48\textwidth]{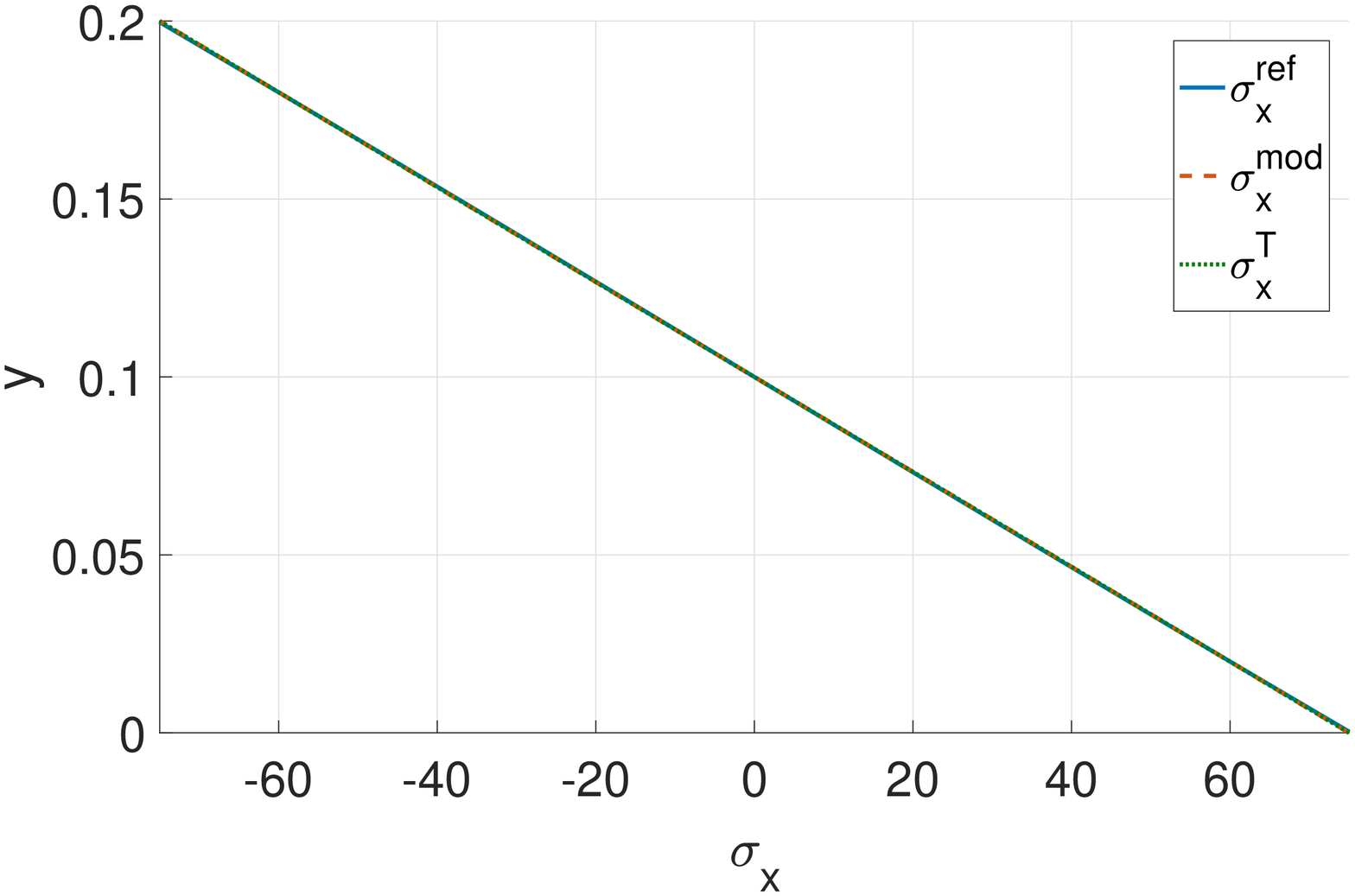}}
\subfigure[shear stress, $x=1$]{
\label{f_homog_tau1}
\includegraphics[width=0.48\textwidth]{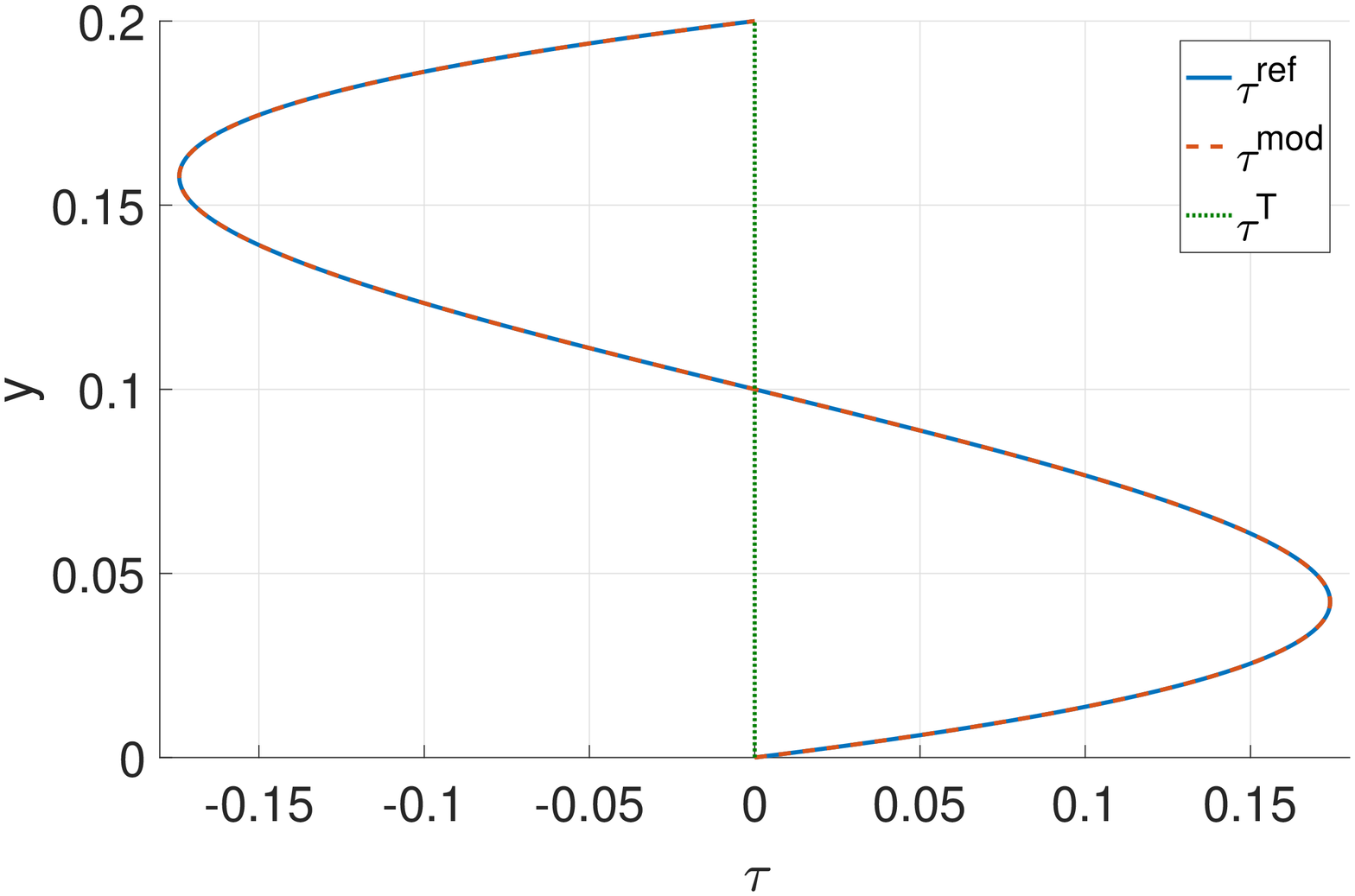}}
\subfigure[axial stress, $x=1.65$]{
\label{f_homog_sx2}
\includegraphics[width=0.48\textwidth]{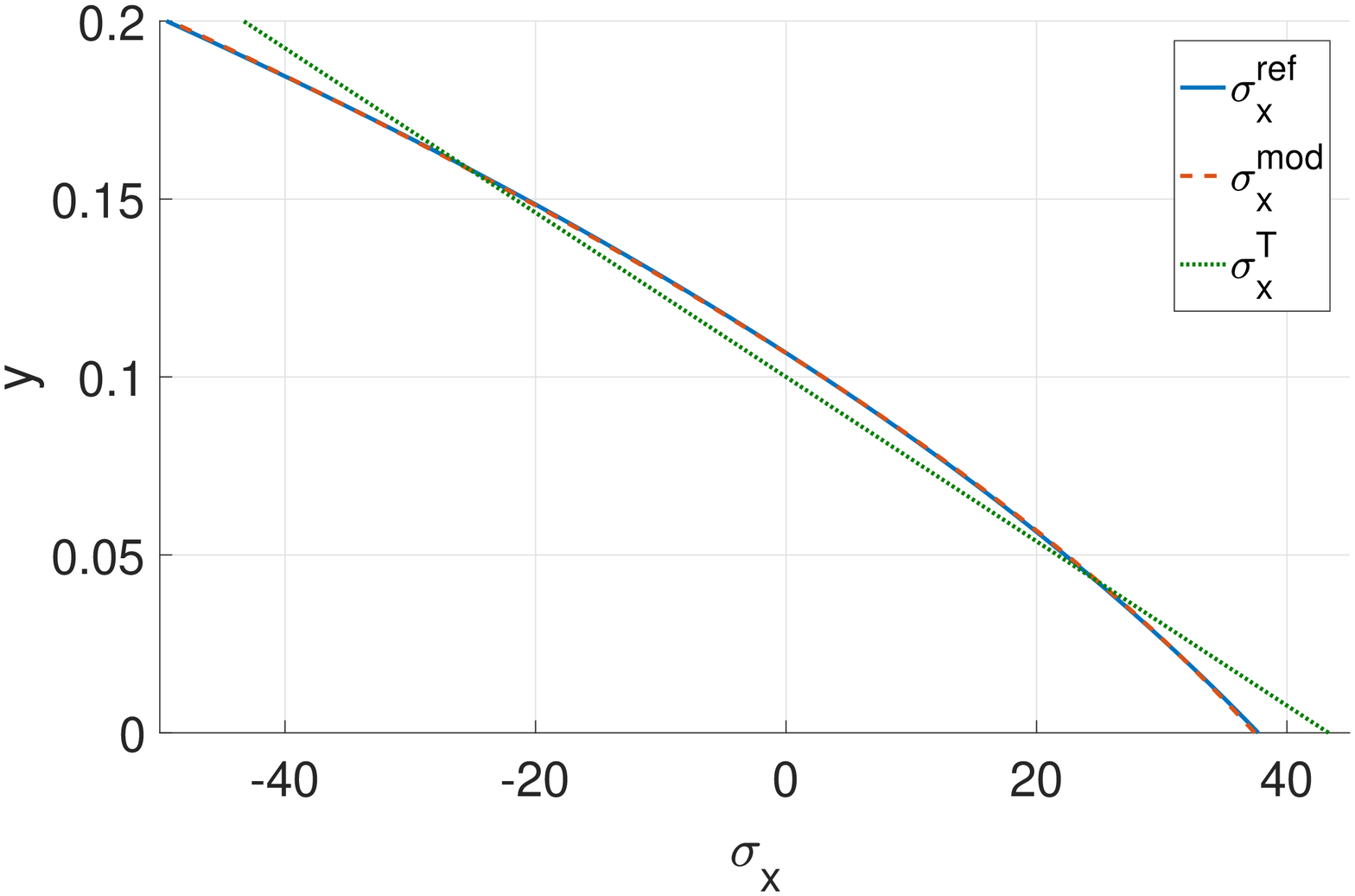}} 
\subfigure[shear stress, $x=1.65$]{
\label{f_homog_tau2}
\includegraphics[width=0.48\textwidth]{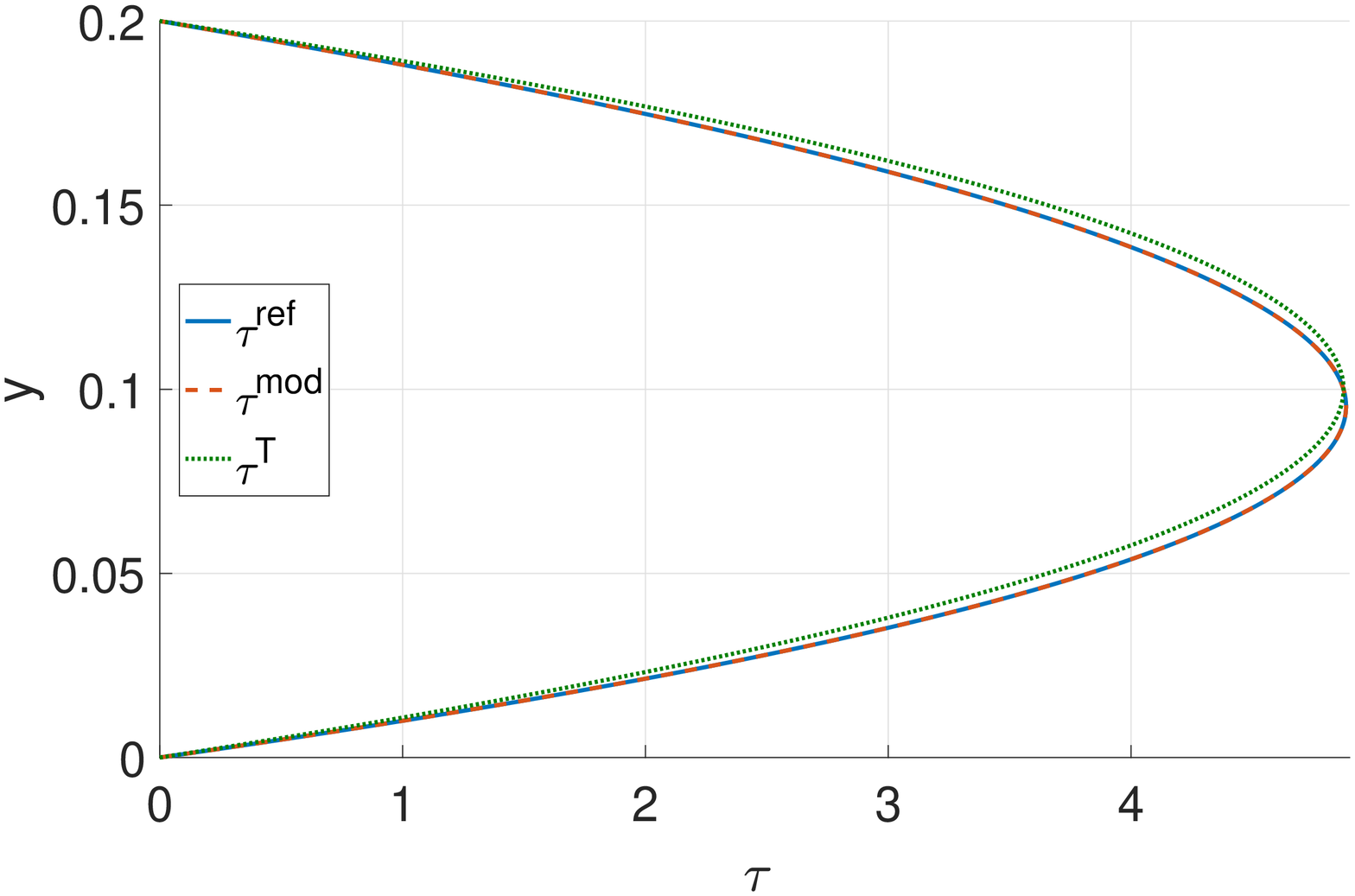}}
\subfigure[axial stress, $x=2$]{
\label{f_homog_sx3}
\includegraphics[width=0.48\textwidth]{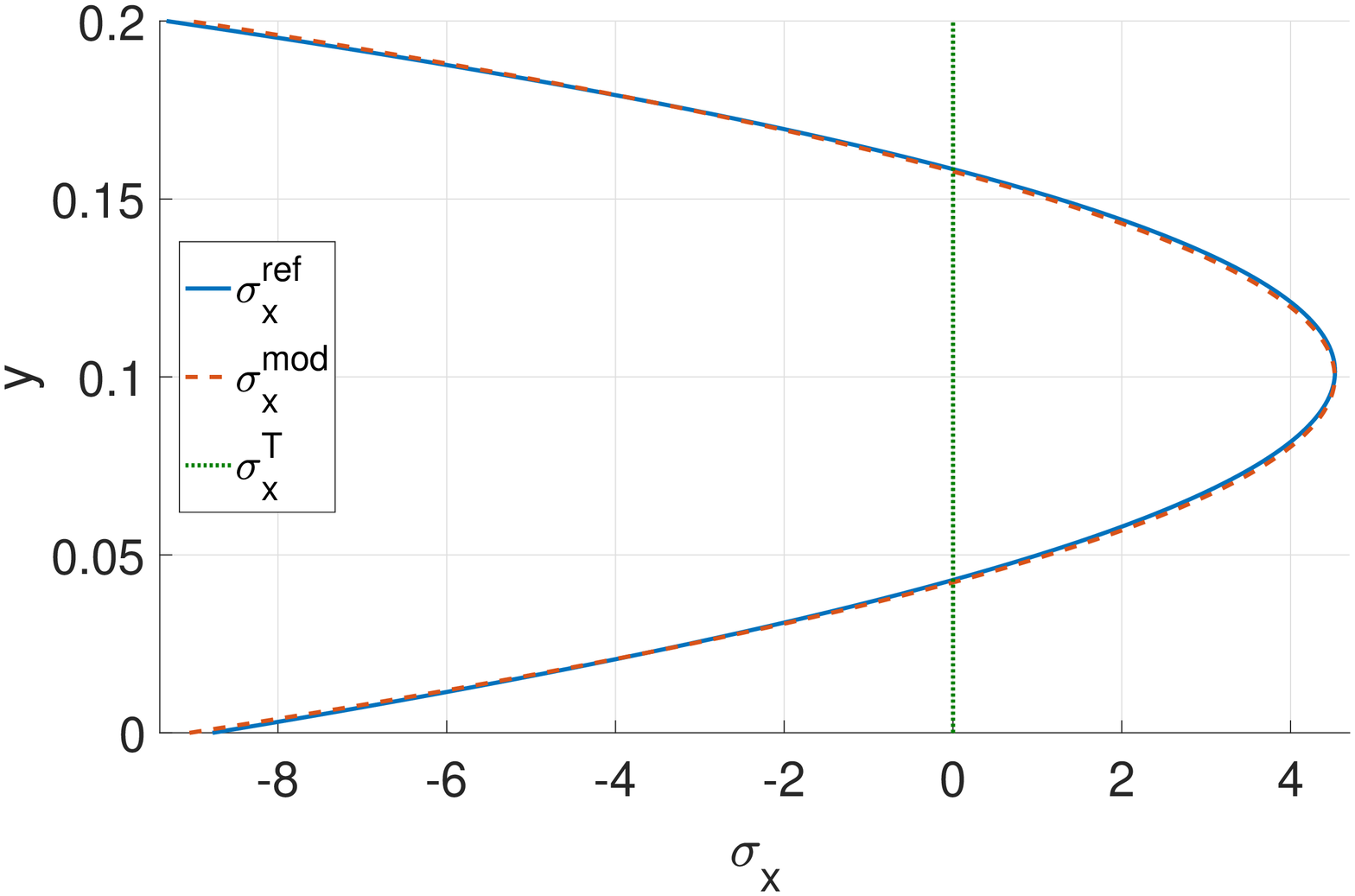}}
\subfigure[shear stress, $x=2$]{
\label{f_homog_tau3}
\includegraphics[width=0.48\textwidth]{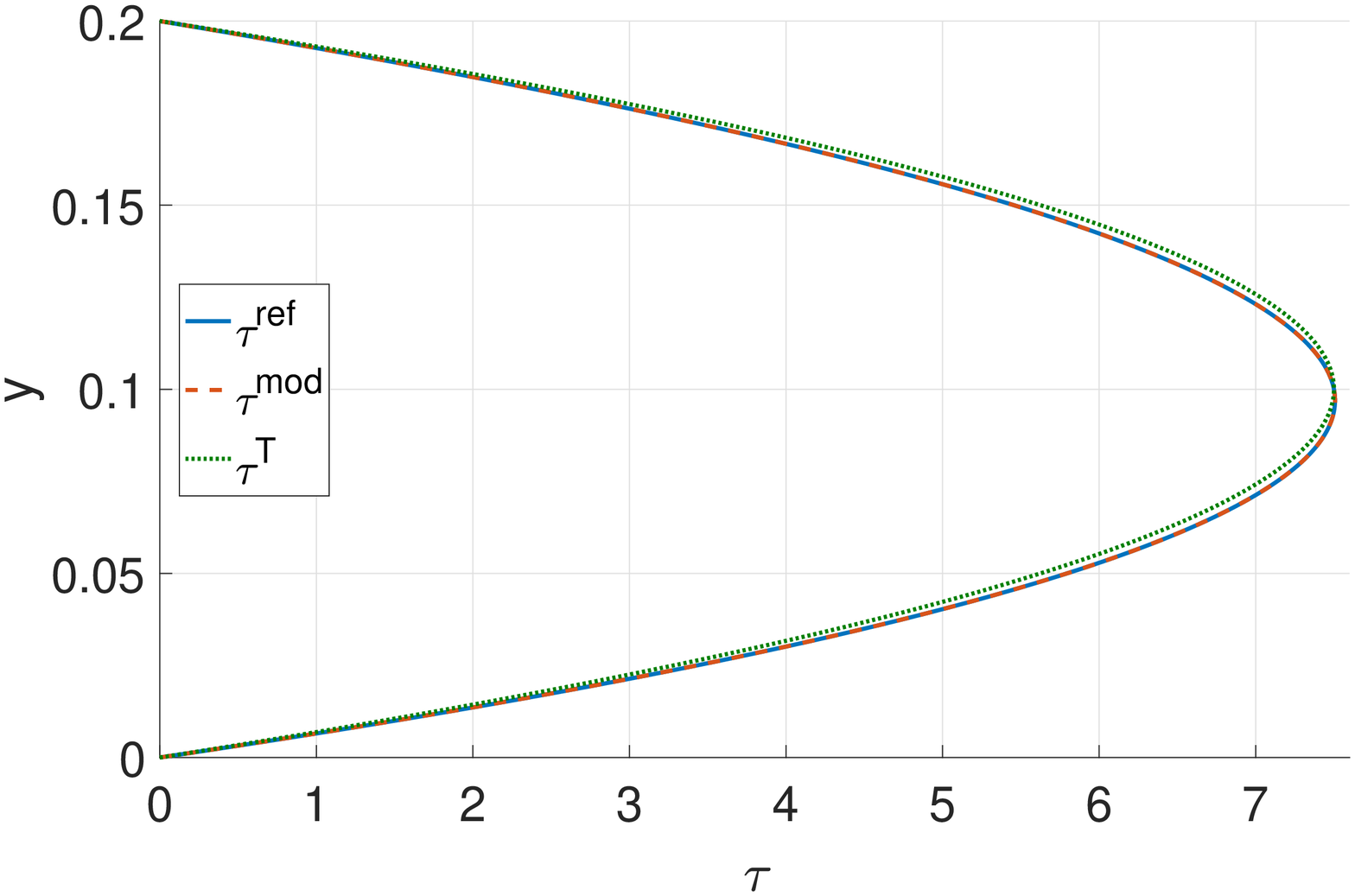}}
\caption{\footnotesize{
Homogeneous, simply-supported, anisotropic beam ($\theta = 45 \, \deg$). 
Analysis of cross-section stress distributions.
Comparisons of reference $\psi^{ref}$, beam model $\psi^{mod}$, and Timoshenko $\psi^{T}$ solutions.}} 
\label{f_homog_stress}
\end{figure}

Figure \ref{f_homog_stress} compares the cross-section distribution of stresses evaluated according to reference \citep{kh_16} $\psi^{ref}$, the proposed beam model $\psi^{mod}$, and standard Timoshenko beam $\psi^{T}$.
Numerical results demonstrate that the proposed beam model provides results substantially identical to the reference solution.
In particular, Figure \ref{f_homog_tau1} highlights that shear does not vanish in the beam mid-span $\tau \left(l/2,y\right) \neq 0 $, despite the vertical internal force vanishes $V \left(l/2\right) = 0$.
Further comments about this  peculiarity of simply supported beams can be found in \citep{kh_16}.
More interestingly, Figure \ref{f_homog_tau1} highlights that axial stress does not vanish at the bearing $\sigma_x \left(l,y\right) \neq 0 $, despite both bending moment and axial internal force vanish $M \left(l\right) = N \left(l\right) = 0$.
Considering also Figures \ref{f_homog_sx2} and \ref{f_homog_tau2}, it is possible to conclude that anisotropy influences the distribution of both axial and shear stresses.
In particular, stress-recovery procedures developed for isotropic structural elements can underestimate the maximal magnitude of axial stress with errors greater than $10 \, \%$.

\section{Comparison with 2D \ac{FE}, bi-layer beam}
\label{s_num_res}

This section reports numerical results for two examples:
Subsection \ref{s_cantilever} considers the cantilever already introduced in Section \ref{s_anal_sol} and Subsection \ref{s_clamp_clamp} analyzes a doubly-clamped beam (see Figure \ref{f_cantilever}).
In both cases, numerical results are obtained assuming the following parameters 
\begin{equation}
\label{mech_prop}
\begin{split}
& h = 100 \, \milli\meter; \quad
q = 1 \, \newton/\milli\meter; \quad
\alpha = 0.5
\\
E_{11} = 10^4 \, \mega & \pascal; \quad
E_{22} = 5 \cdot10^2 \, \mega\pascal; \quad
G = 10^3 \, \mega\pascal; \quad
\nu = 0
\end{split}
\end{equation}

In this section, the reference solution $\psi^{ref}$ is computed using the commercial software Abaqus \citep{abaqus}, in which the 2D problem domain $\Omega$ was discretized with a structured mesh of square bilinear elements CPS4.
As discussed in Section \ref{s_intro}, boundary effects could significantly affect the structural element behavior.
Aiming at limiting their influence in reference solution, the \acp{BC} are imposed requiring only vanishing mean value of cross-section displacements and rotation.
In this manner constrained cross-sections can warp and deform, but stress concentrations are limited in magnitude.
Aiming at guaranteeing negligible numerical errors in the reference results, a sequence of analysis has been performed considering the bilayer cantilever and defining the element size $\delta$ according to the series $1/2^n$ for $n=0,1,2, \dots$.
The procedure has been interrupted when the relative increase of the maximal displacement magnitude was smaller than $10^{-4}$, leading to set $\delta = 0.25$

Transversal displacement $v^{ref} \left( x \right)$ and shear strain $\gamma^{ref} \left( x \right)$ have been obtained computing the mean value over the cross-section of the 2D transversal displacements $s_y^{ref} \left(x,y\right)$ and shear strains $\gamma_{xy}^{ref} \left(x,y\right)$, respectively.
Conversely, axial $N^{ref} \left( x \right)$ and shear $V^{ref} \left( x \right)$ internal forces have been obtained as the integral over the cross-section of stress components $\sigma_{x}^{ref} \left(x,y\right)$ and $\tau^{ref} \left(x,y\right)$, respectively.
The bending moment $M^{ref} \left( x \right)$ has been obtained as the integral over the cross section of axial stress $\sigma_{x}^{ref} \left(x,y\right)$ times the $y$ coordinate.
Finally, according to Remark \ref{r_approx}, the axial displacement $u ^{ref} \left( x \right)$ and the rotation $\phi^{ref} \left( x \right)$ have been computed as the coefficients of the linear least squares with respect to $y$ of the axial displacements $s_x^{ref} \left(x,y\right)$.
Similarly, the axial strain $\epsilon ^{ref} \left( x \right)$ and the curvature $\chi^{ref} \left( x \right)$ have been computed as the coefficients of the linear least squares with respect to $y$ of the strains $\epsilon_x^{ref} \left(x,y\right)$.


\subsection{Cantilever}
\label{s_cantilever}

In the following we set $l = 500 \, \milli\meter$ i.e., we choose $\lambda = 5$.
This assumption leads to consider a beam geometry that is close to the well known limit of validity of the \acp{FSDT} and, therefore, it will allow to identify every potential critical issue of the proposed model.
Assuming $\theta = 15 \, \deg$ the rotation of constitutive relation \eqref{const_rel_rotation} leads to set
\begin{equation}
\label{mech_prop1}
\mu = 1.5853; \quad
\kappa = 1.2750; \quad
G_x = -4.2222 \cdot 10^3 \, \mega \pascal
\end{equation}

\begin{figure}[htbp]
\centering
\subfigure[axial displacement]{
\label{f_layer+_u}
\includegraphics[width=0.48\textwidth]{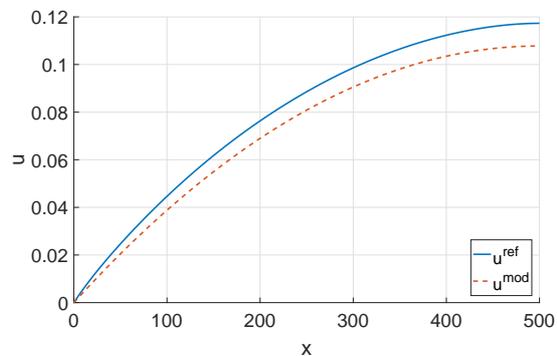}}\\
\subfigure[rotation components]{
\label{f_layer+_fp}
\includegraphics[width=0.48\textwidth]{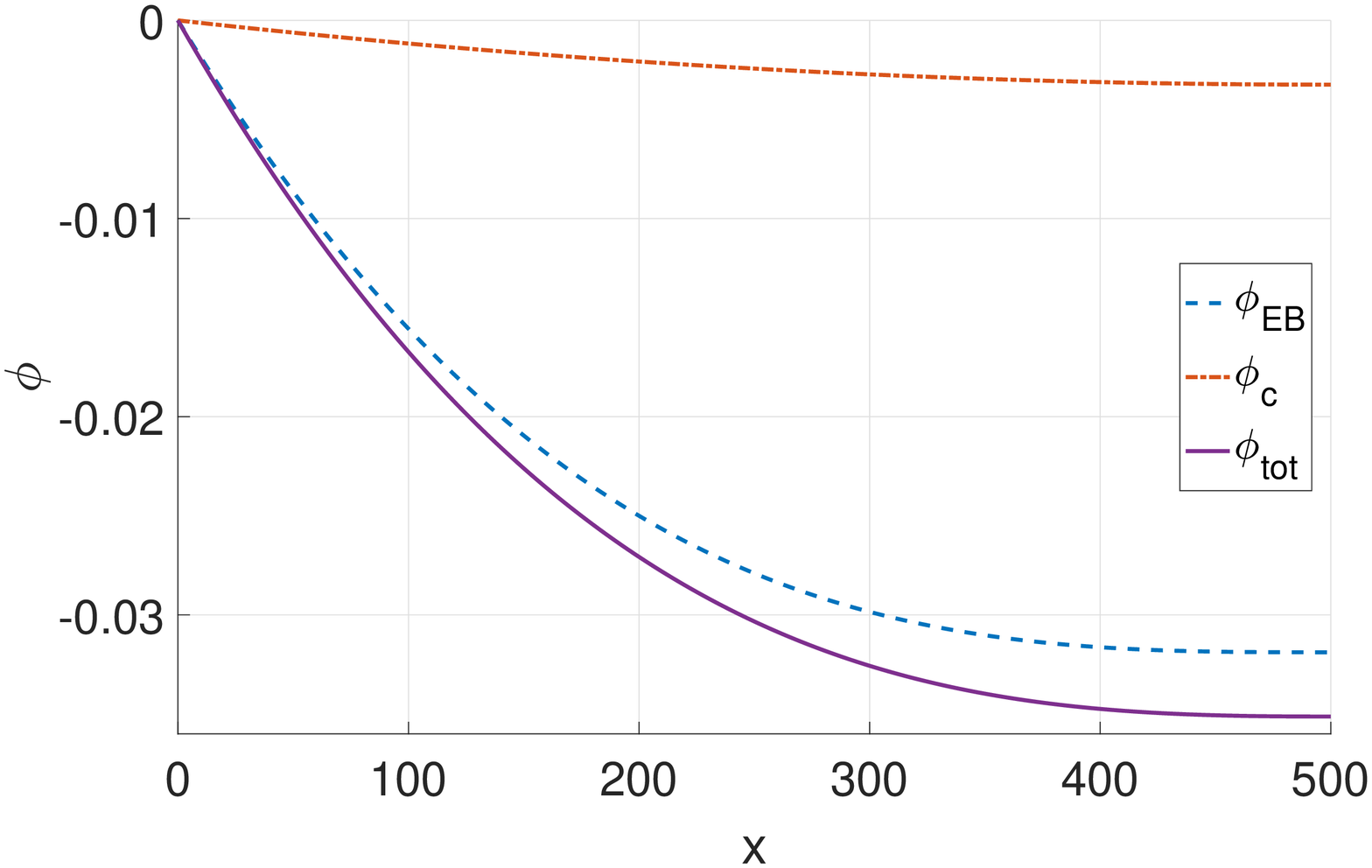}}
\subfigure[rotation]{
\label{f_layer+_f}
\includegraphics[width=0.48\textwidth]{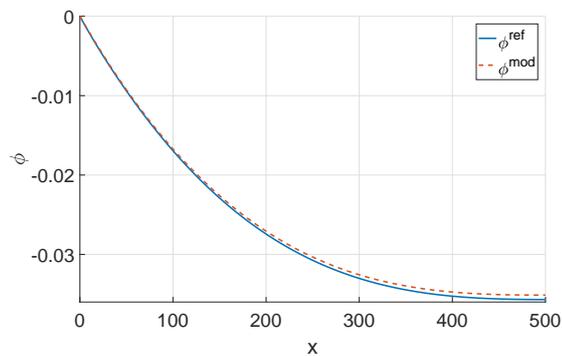}} 
\subfigure[transversal displacement components]{
\label{f_layer+_vp}
\includegraphics[width=0.48\textwidth]{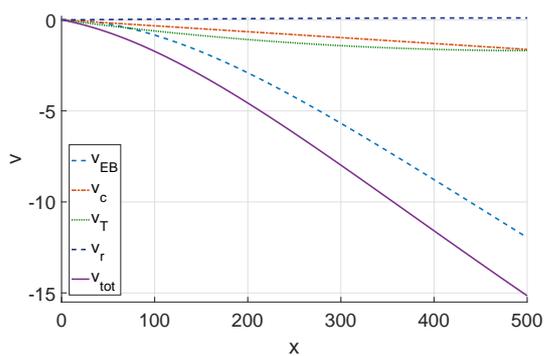}}
\subfigure[transversal displacement]{
\label{f_layer+_v}
\includegraphics[width=0.48\textwidth]{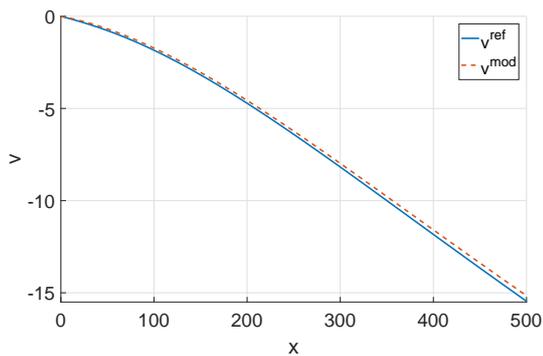}}
\caption{\footnotesize{
Bi-layer anisotropic cantilever ($\theta = 15 \, \deg$). 
Analysis of the generalized displacements components according to the proposed beam model (Figures \ref{f_layer+_fp} and \ref{f_layer+_vp}).
Comparisons of the beam model $\psi^{mod}$ and the reference $\psi^{ref}$ solutions (Figures \ref{f_layer+_u}, \ref{f_layer+_f}, and \ref{f_layer+_v}).}} 
\label{f_layer+_disp}
\end{figure}

Figure \ref{f_layer+_disp} reports numerical results concerning the displacements.
Figures \ref{f_layer+_fp} and \ref{f_layer+_vp} analyze the model solution, highlighting the deep influence of the material coupling term $G_{x}$ on global structural response.
In particular, Figure \ref{f_layer+_vp} highlights that transversal displacement component $v_{c} \left(x\right)$ (see Equation \eqref{ode_sol}) has a magnitude similar to the shear deformation component $v_{T} \left(x\right)$ and it contributes to total transversal displacement more than $10 \, \%$.
Conversely, $v_{r} \left(x\right)$ influences the total displacement less than $1 \, \%$.
Figure \ref{f_layer+_fp} highlights that $\phi_{c} \left(x\right)$ contributes to total cross-section rotation up to $10 \, \%$.
Figures \ref{f_layer+_f} and \ref{f_layer+_v} solution reveal a good accuracy of the proposed model. 
Indeed, relative errors are smaller than $2 \, \%$ for rotation and transversal displacements.

Due to considered loads $N^{mod} \left(x\right) = 0$ and, due to \acp{BC}, also $u_{EB} \left(x\right) = 0$ (see Equation \eqref{ode_sol}).
As a consequence, $u^{mod} \left(x\right) = u_{c} \left(x\right)$ i.e., the axial displacement is uniquely controlled by the material coupling term $G_{x}$ and the transversal internal force $V \left(x\right)$.
Figure \ref{f_layer+_u} shows that beam model correctly predict a non-vanishing distribution of axial displacement, but the error is near to $8 \, \%$.
Nevertheless, the axial displacement is two order of magnitude smaller than the transversal one, not affecting the errors evaluated on the total displacement $\pmb{s} \left(x,y\right)$.

\begin{figure}[htbp]
\centering
\subfigure[axial strain components]{
\label{f_theta+_ep}
\includegraphics[width=0.48\textwidth]{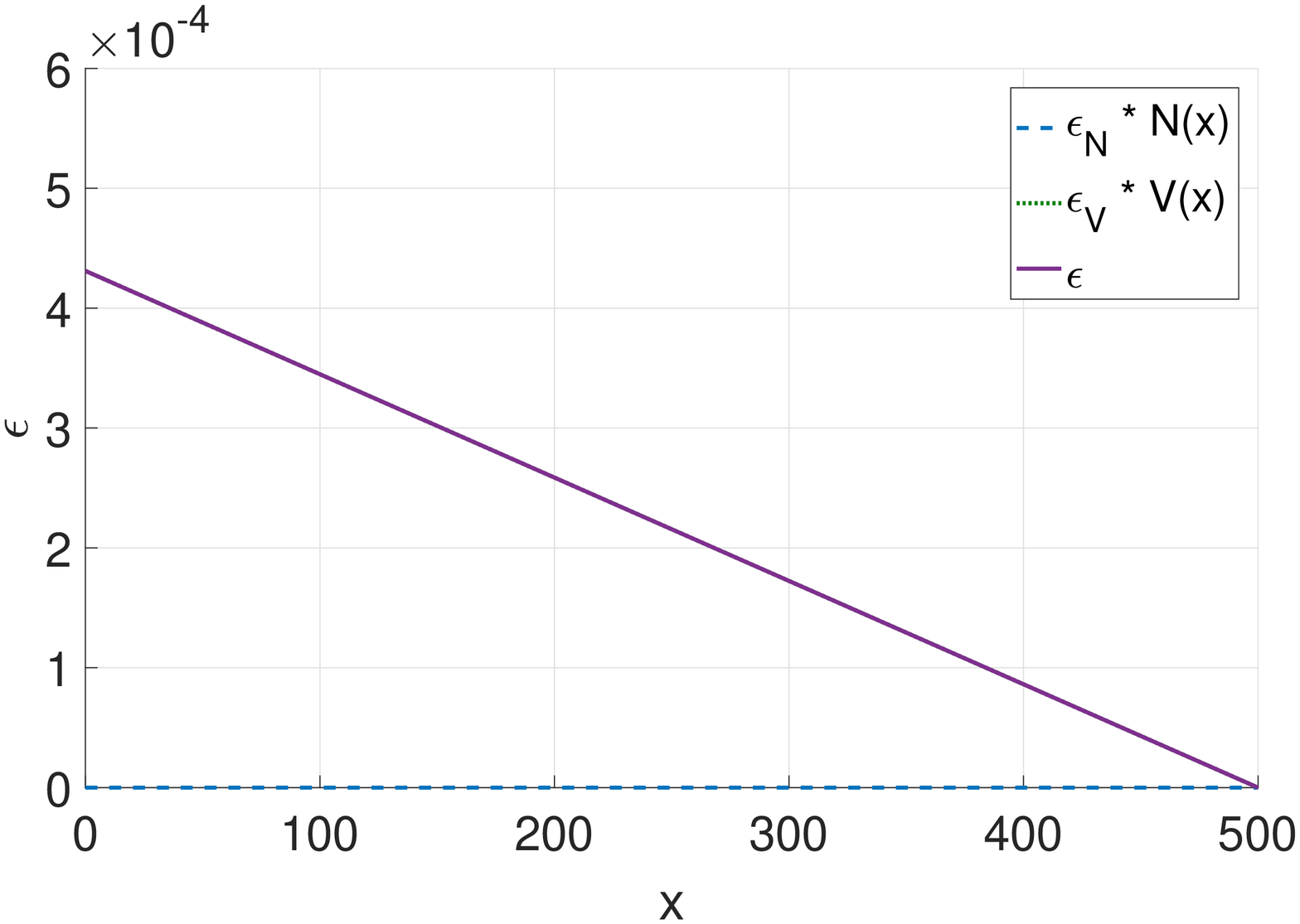}}
\subfigure[axial strain]{
\label{f_theta+_e}
\includegraphics[width=0.48\textwidth]{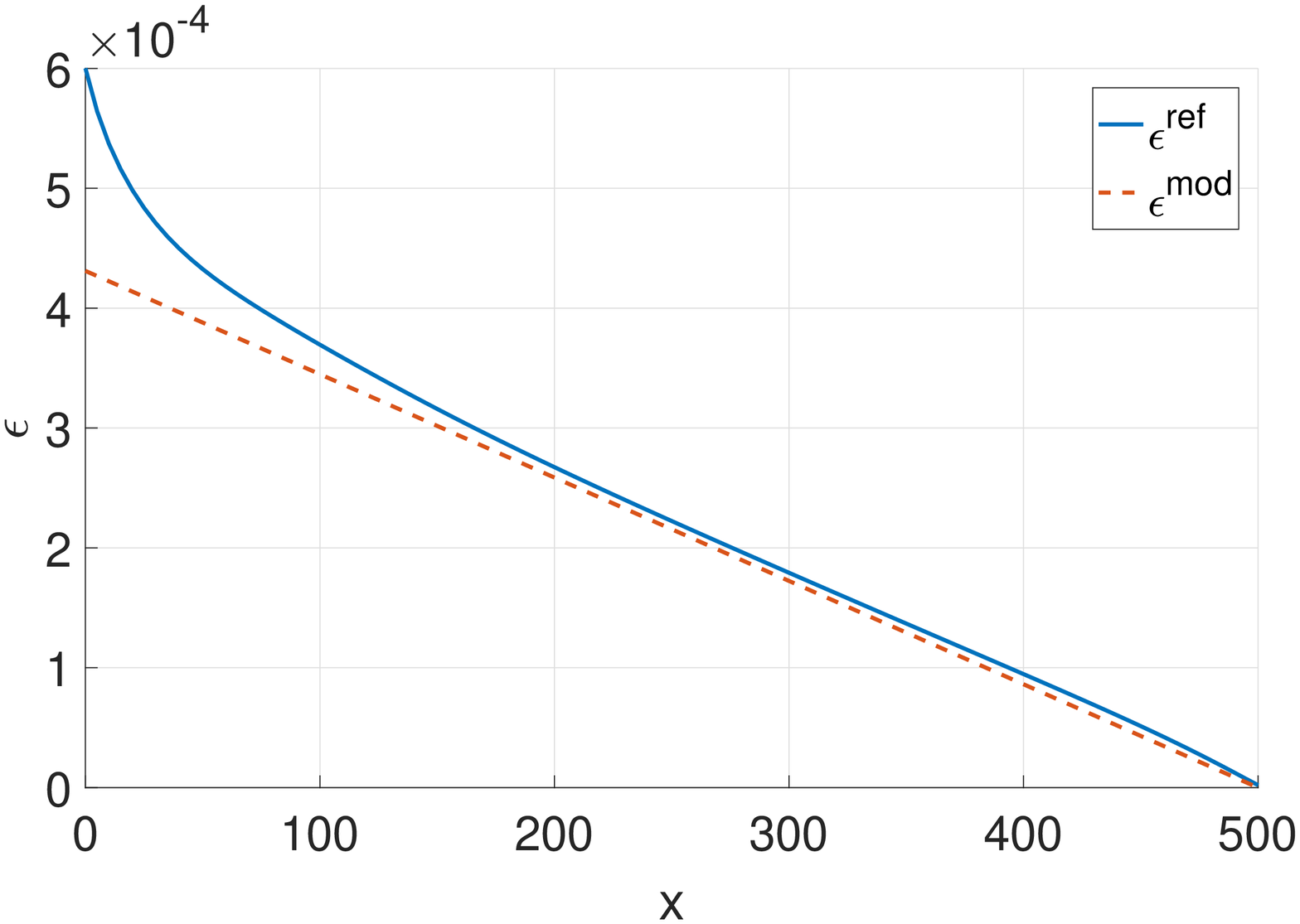}}
\subfigure[curvature components]{
\label{f_theta+_cp}
\includegraphics[width=0.48\textwidth]{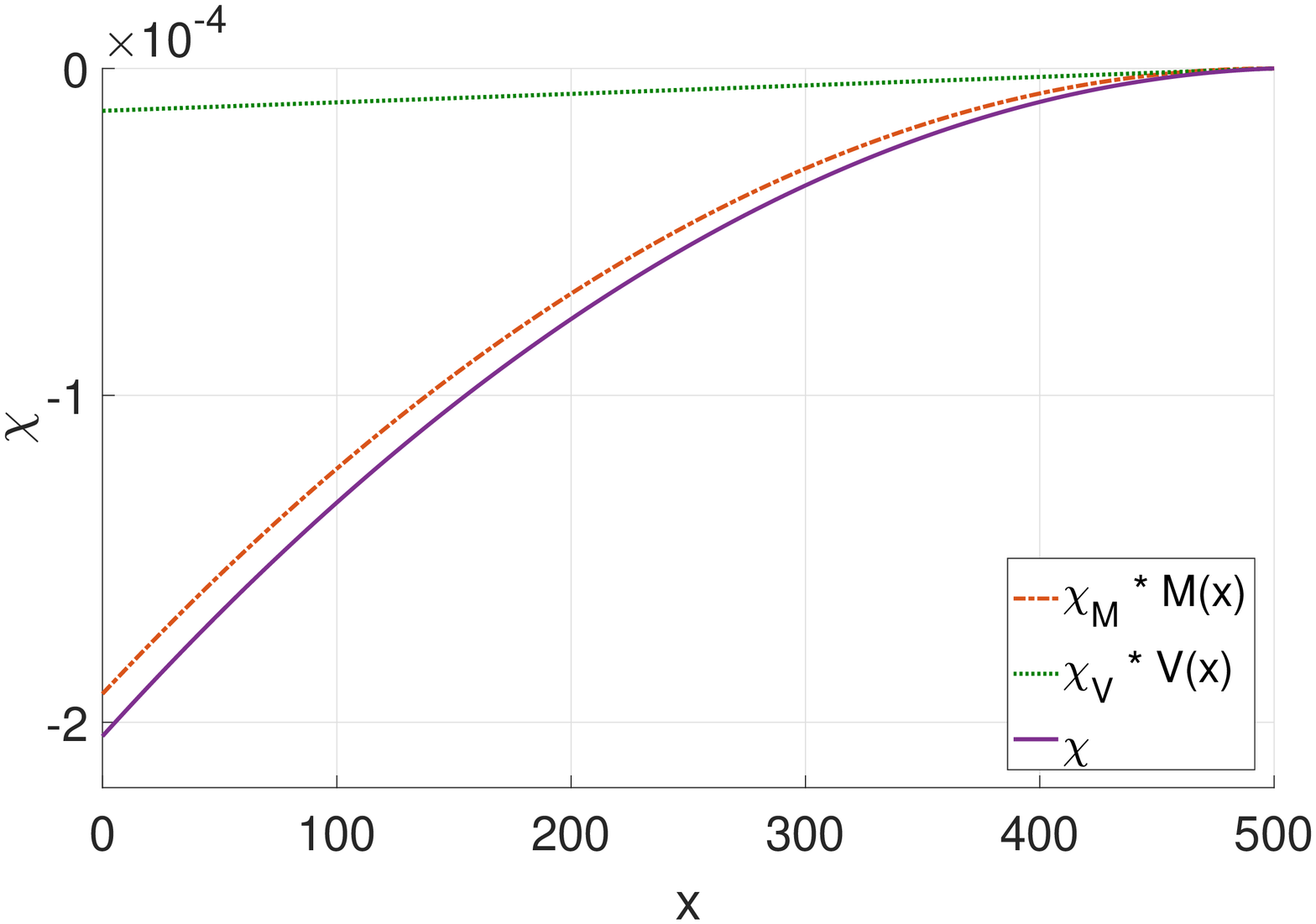}}
\subfigure[curvature]{
\label{f_theta+_c}
\includegraphics[width=0.48\textwidth]{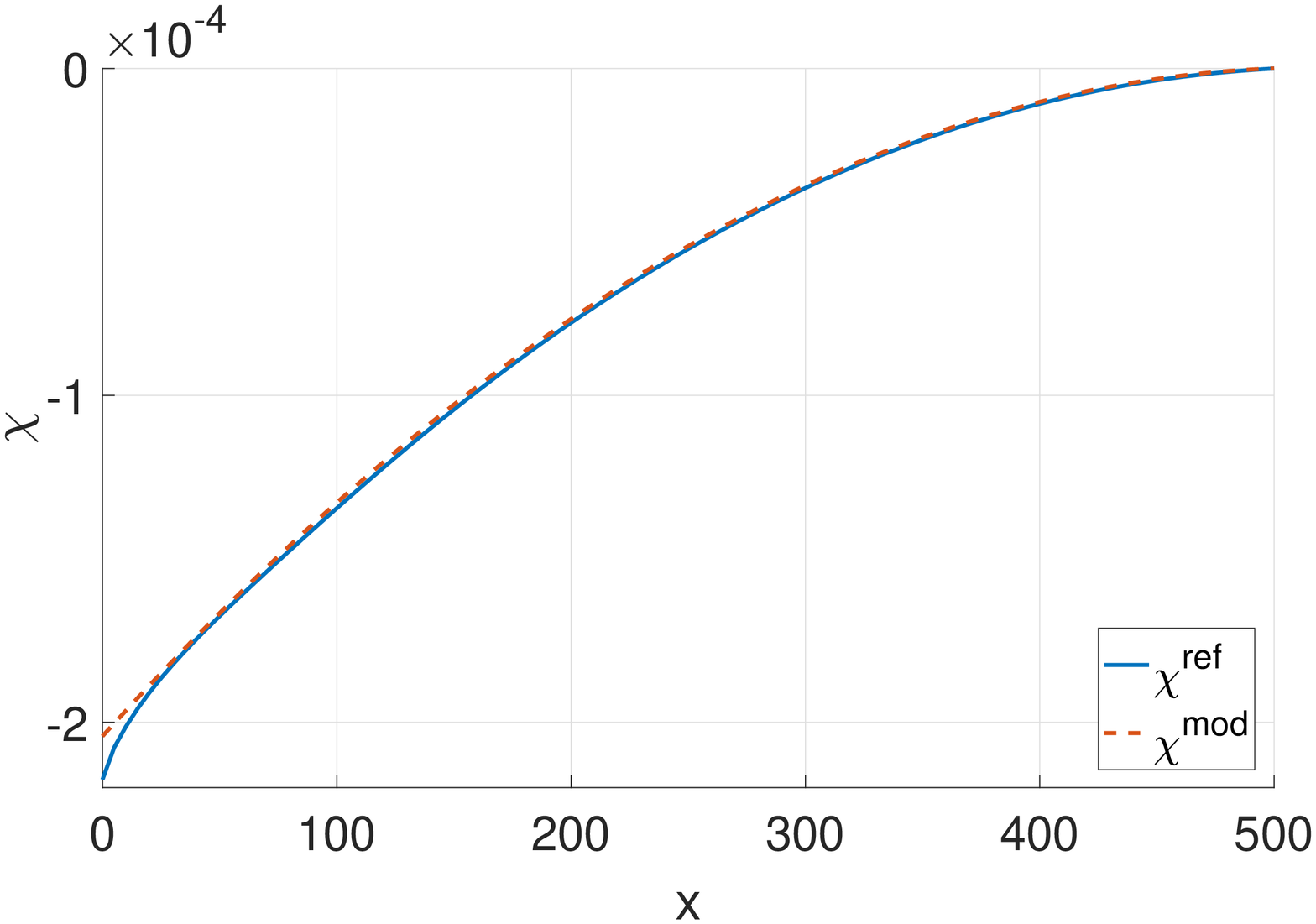}}
\subfigure[shear strain components]{
\label{f_theta+_gp}
\includegraphics[width=0.48\textwidth]{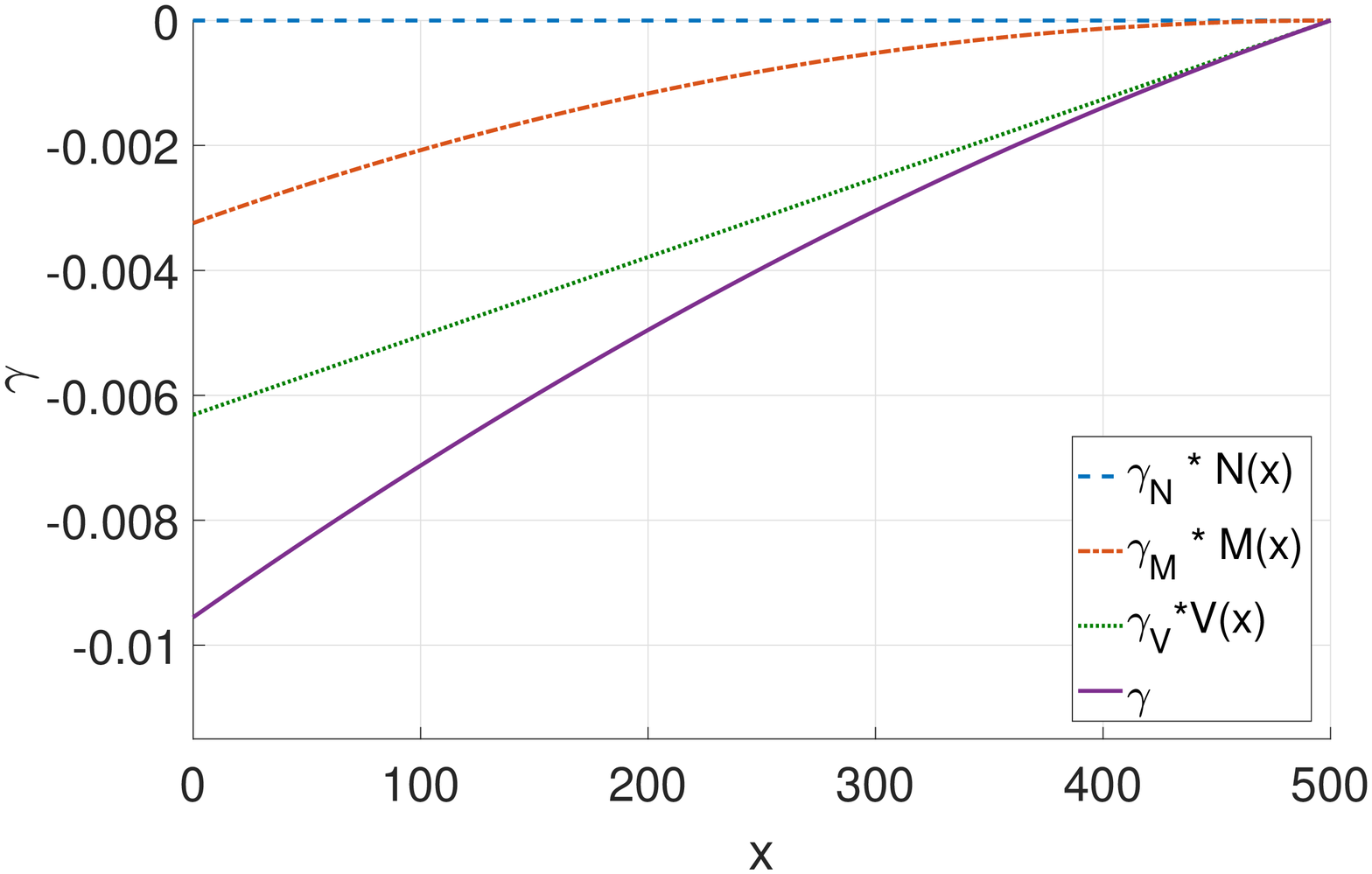}}
\subfigure[shear strain]{
\label{f_theta+_g}
\includegraphics[width=0.48\textwidth]{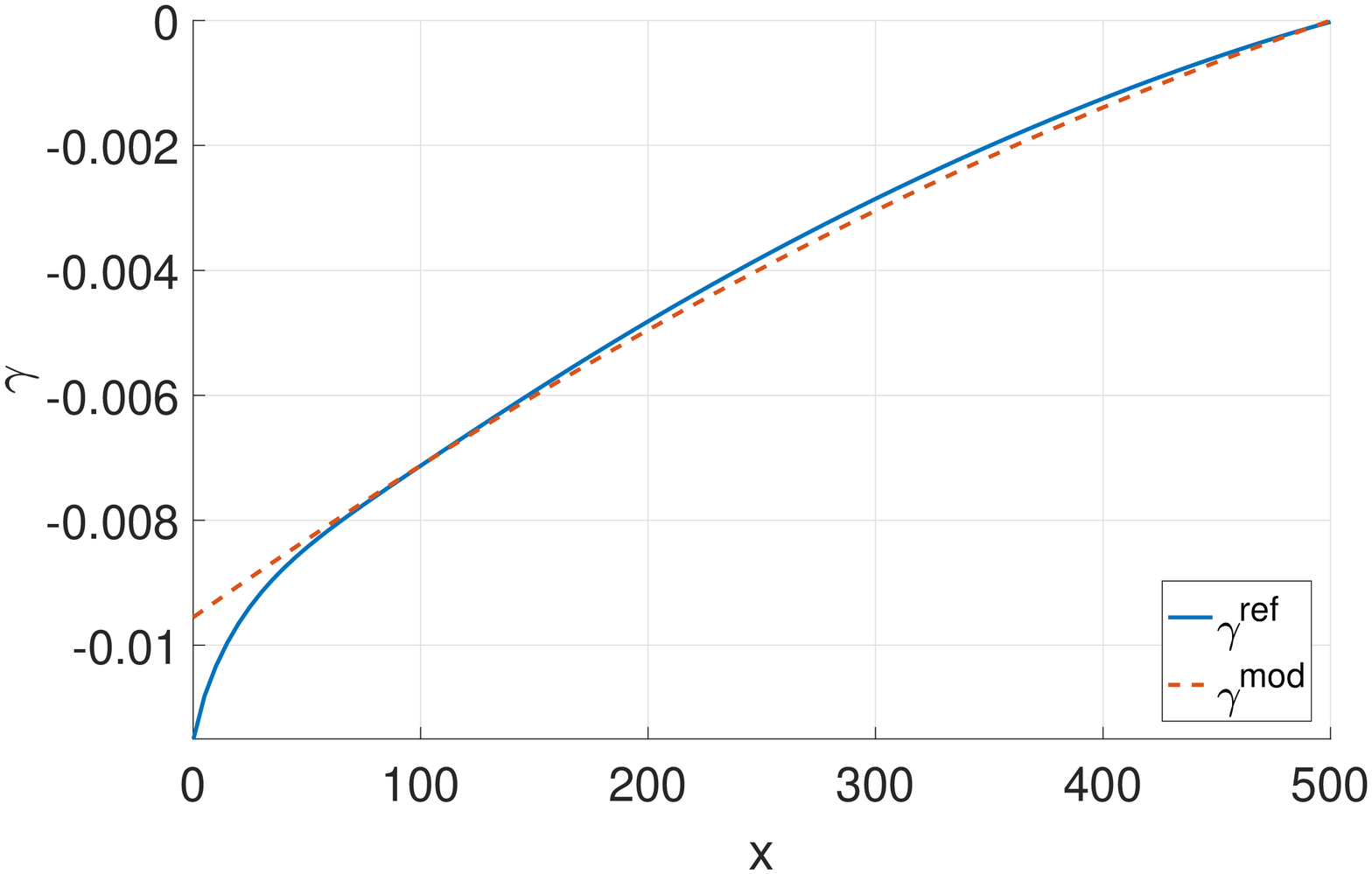}}
\caption{\footnotesize{
Bi-layer anisotropic cantilever ($\theta = 15 \, \deg$). 
Analysis of the generalized strains components according to the proposed beam model (Figures \ref{f_theta+_ep}, \ref{f_theta+_cp}, and \ref{f_theta+_gp}). 
Comparisons of the beam model $\psi^{mod}$ and the reference $\psi^{ref}$ solutions (Figures \ref{f_theta+_e}, \ref{f_theta+_c}, and \ref{f_theta+_g}).}} 
\label{f_theta+_defo}
\end{figure}

Figure \ref{f_theta+_defo} reports numerical results concerning generalized strains.
Figure \ref{f_theta+_ep} shows that axial strain is uniquely attributed to the transversal internal force $V \left(x\right)$ by means of the coefficient $\epsilon_V$ as already discussed above (see also Figure \ref{f_layer+_u}).
The comparison with reference solution (Figure \ref{f_theta+_e}) confirms the goodness of the estimation provided by the beam model.
Anyway, the reference solution reveals the presence of some higher order effects close to the clamp that can not be detected by the proposed beam model (see Remark \ref{r_bondary_effects}) and may be also responsible of the errors on axial displacements.

Figure \ref{f_theta+_cp} shows that transversal internal force $V \left(x\right)$ produces non-negligible curvature, up to $10 \, \%$ of the total.
Similarly, Figure \ref{f_theta+_gp} shows that bending moment $M \left(x\right)$ deeply influences the shear strain which has a non-linear distribution despite transversal internal force $V\left(x\right)$ is linear. In particular, bending moment $M \left(x\right)$ produces non-negligible shear strain, up to $30 \, \%$ of the total.
For both shear deformation and curvature, Figures \ref{f_theta+_c} and \ref{f_theta+_g} demonstrate that generalized strains predicted by the beam model are in extremely good agreement with reference solution.
Only near to the clamp, reference solution reveals the presence of some higher order effects that are not handled by the beam model.

\begin{figure}[htbp]
\centering 
%
%
\subfigure[axial stress components, $x= 250$]{
\label{f_layer+_sx2_par}
\includegraphics[width=0.48\textwidth]{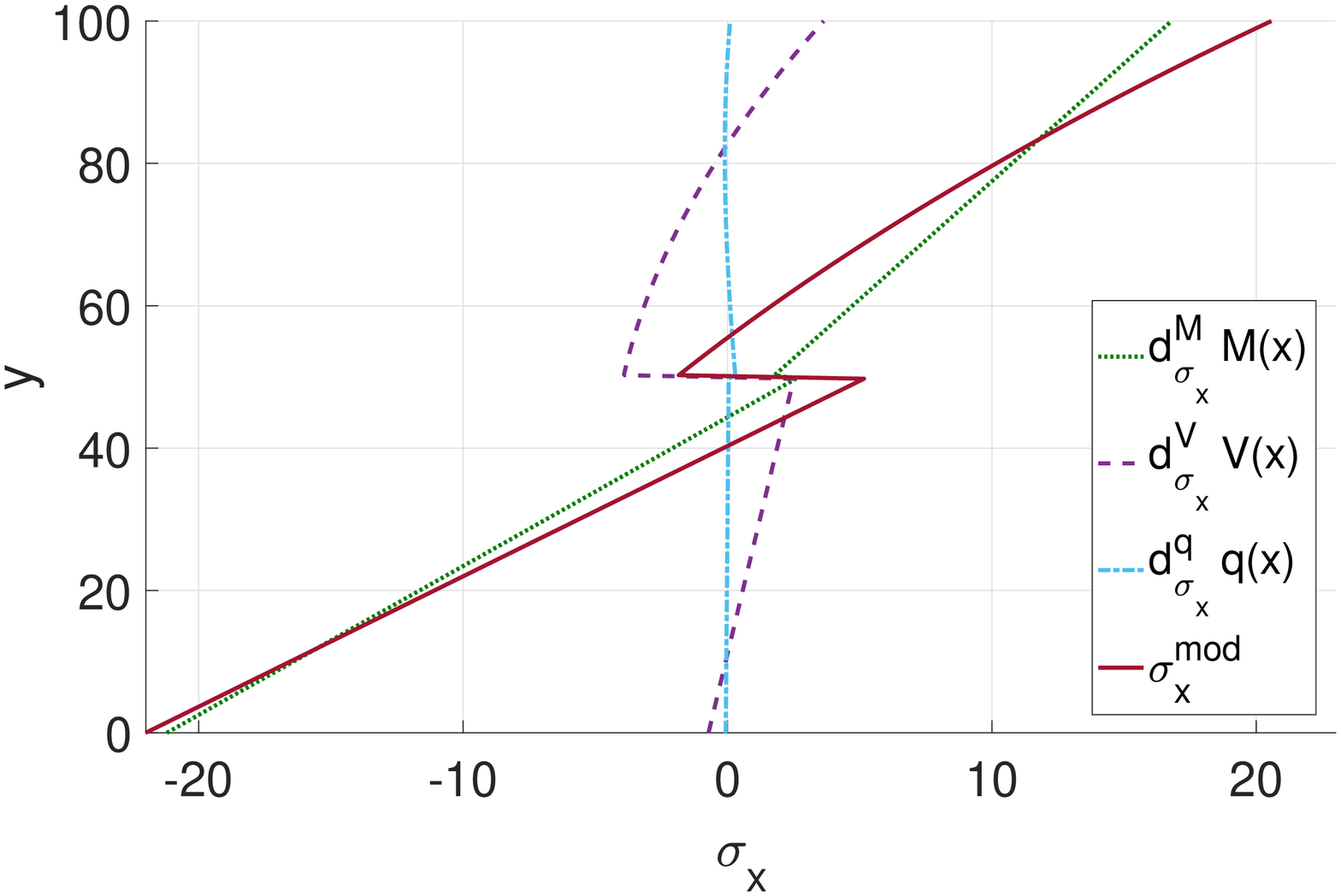}}
\subfigure[axial stress, $x= 250$]{
\label{f_layer+_sx2}
\includegraphics[width=0.48\textwidth]{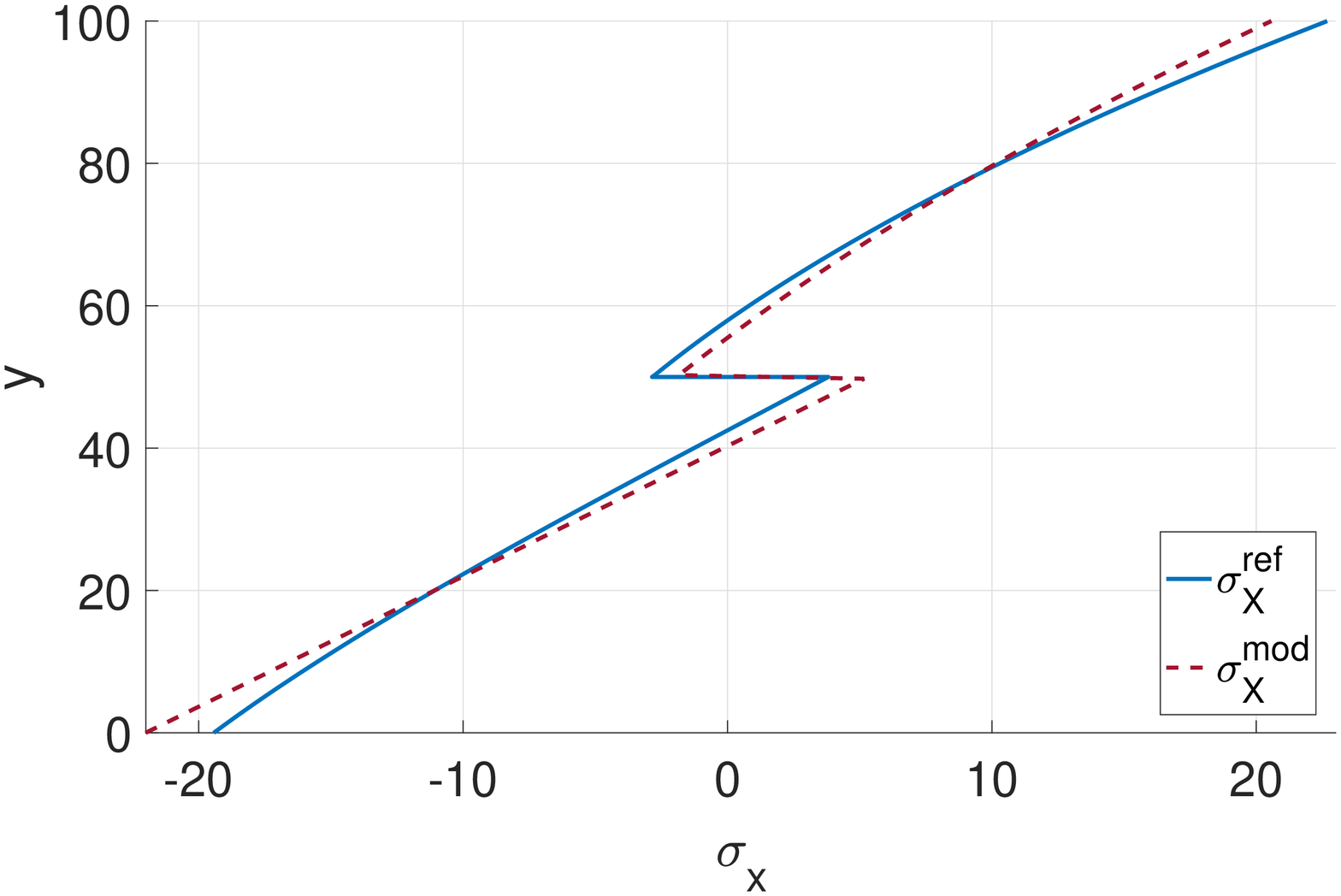}}
\subfigure[axial stress components, $x= 375$]{
\label{f_layer+_sx3_par}
\includegraphics[width=0.48\textwidth]{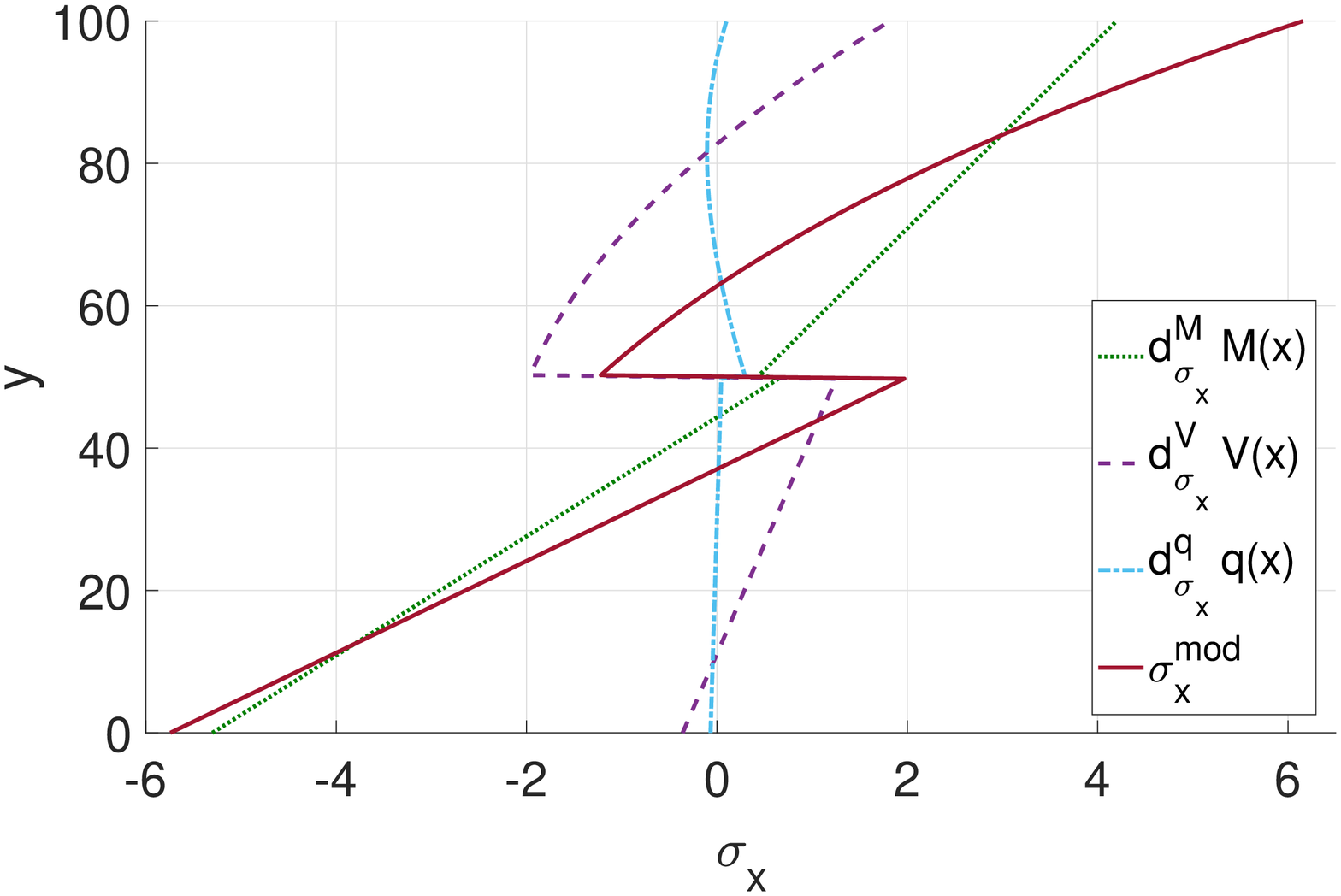}}
\subfigure[axial stress, $x= 375$]{
\label{f_layer+_sx3}
\includegraphics[width=0.48\textwidth]{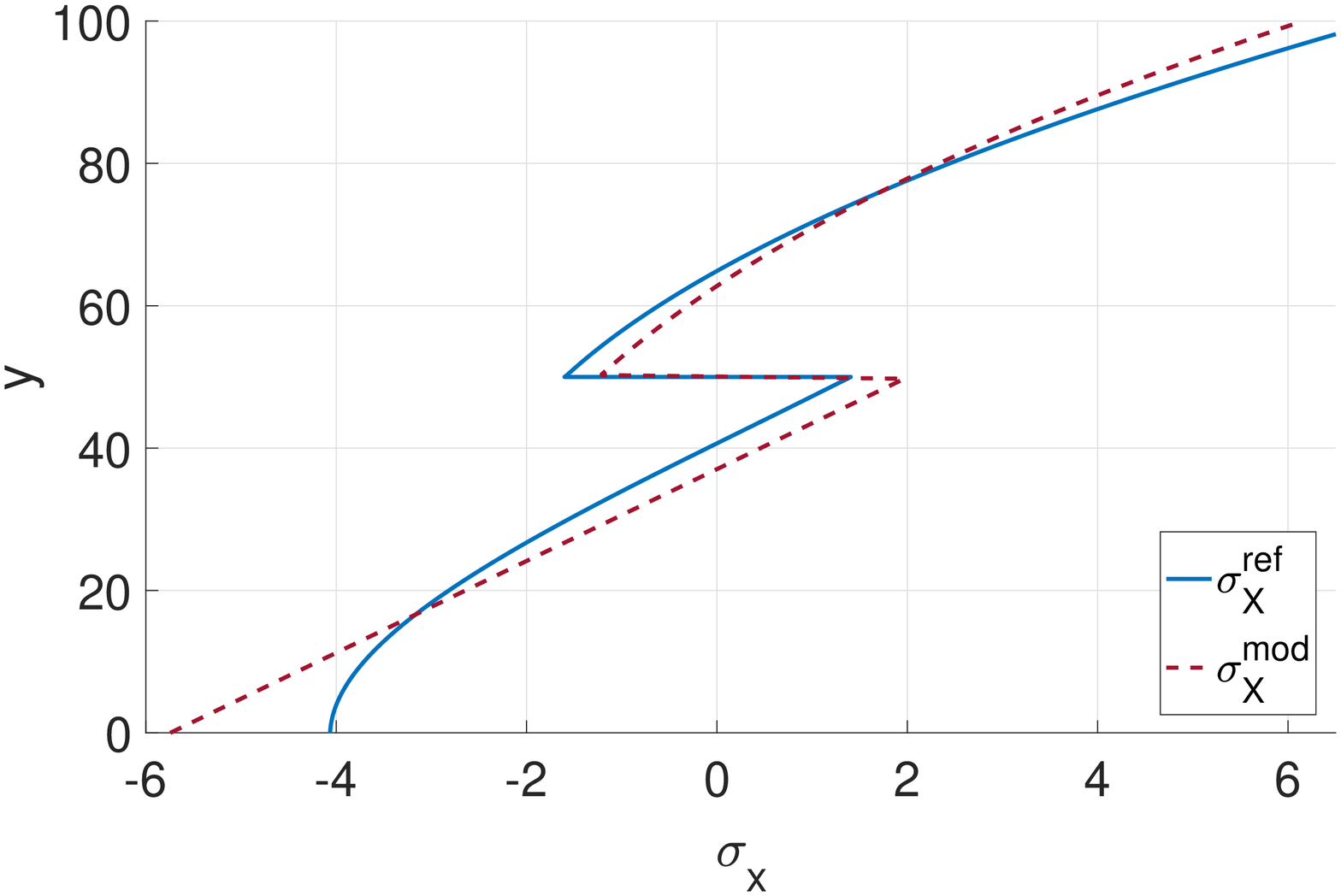}}
\caption{\footnotesize{
Bi-layer anisotropic cantilever ($\theta = 15 \, \deg$).
Axial stress distributions evaluated at $x=250 \, \milli\meter$ (Figures \ref{f_layer+_sx2_par} and \ref{f_layer+_sx2}), and $x=375 \, \milli\meter$ (Figures \ref{f_layer+_sx3_par} and \ref{f_layer+_sx3}).
Analysis of the axial stress components according to the proposed beam model (Figures \ref{f_layer+_sx2_par} and \ref{f_layer+_sx3_par}) and comparisons of the beam model $\psi^{mod}$ and the reference $\psi^{ref}$ solutions (Figures \ref{f_layer+_sx2} and\ref{f_layer+_sx3}).}} 
\label{f_layer+_sx}
\end{figure}

\begin{figure}[htbp]
\centering 
%
%
\subfigure[shear stress components, $x= 250$]{
\label{f_layer+_t2_par}
\includegraphics[width=0.48\textwidth]{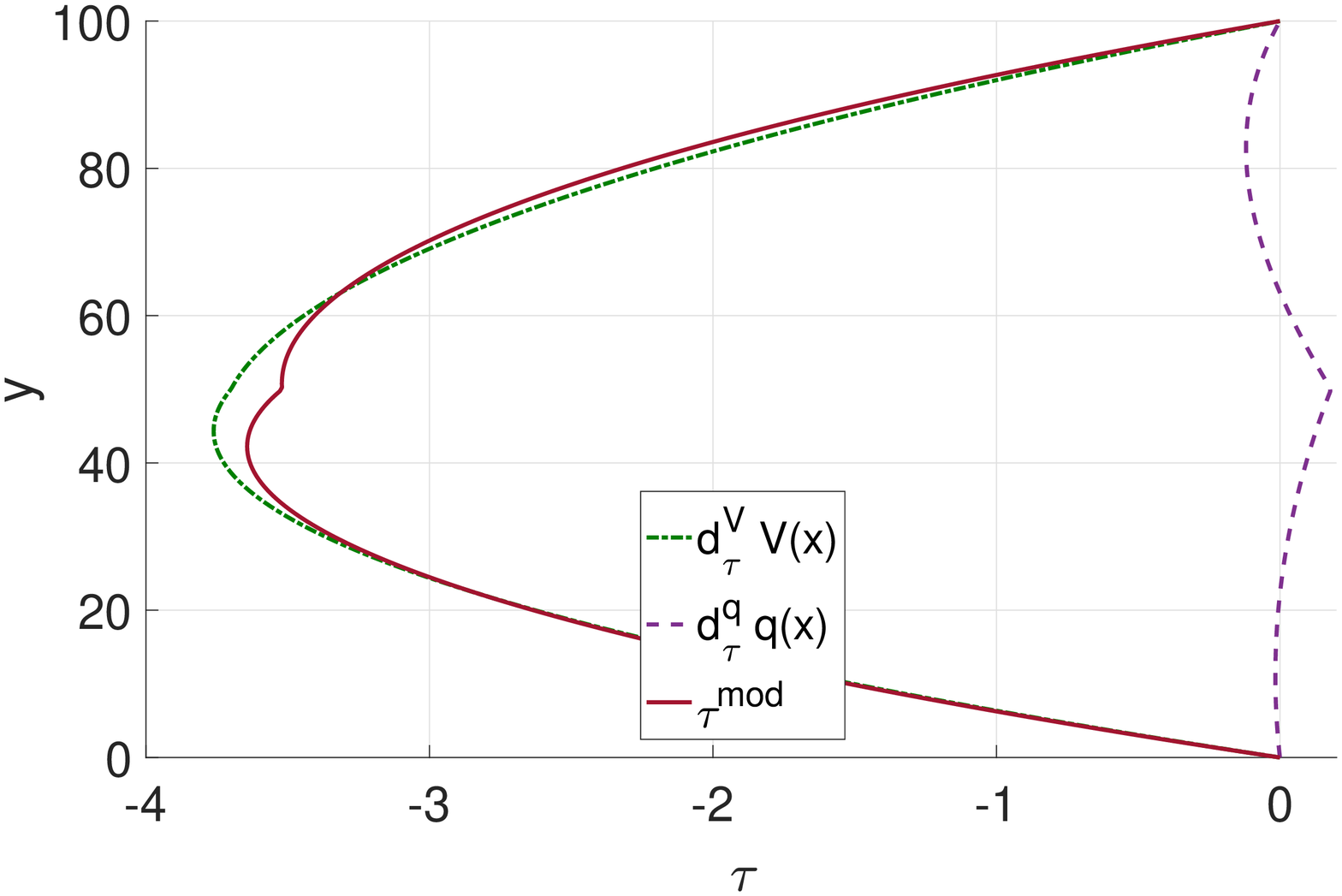}}
\subfigure[shear stress, $x= 250$]{
\label{f_layer+_t2}
\includegraphics[width=0.48\textwidth]{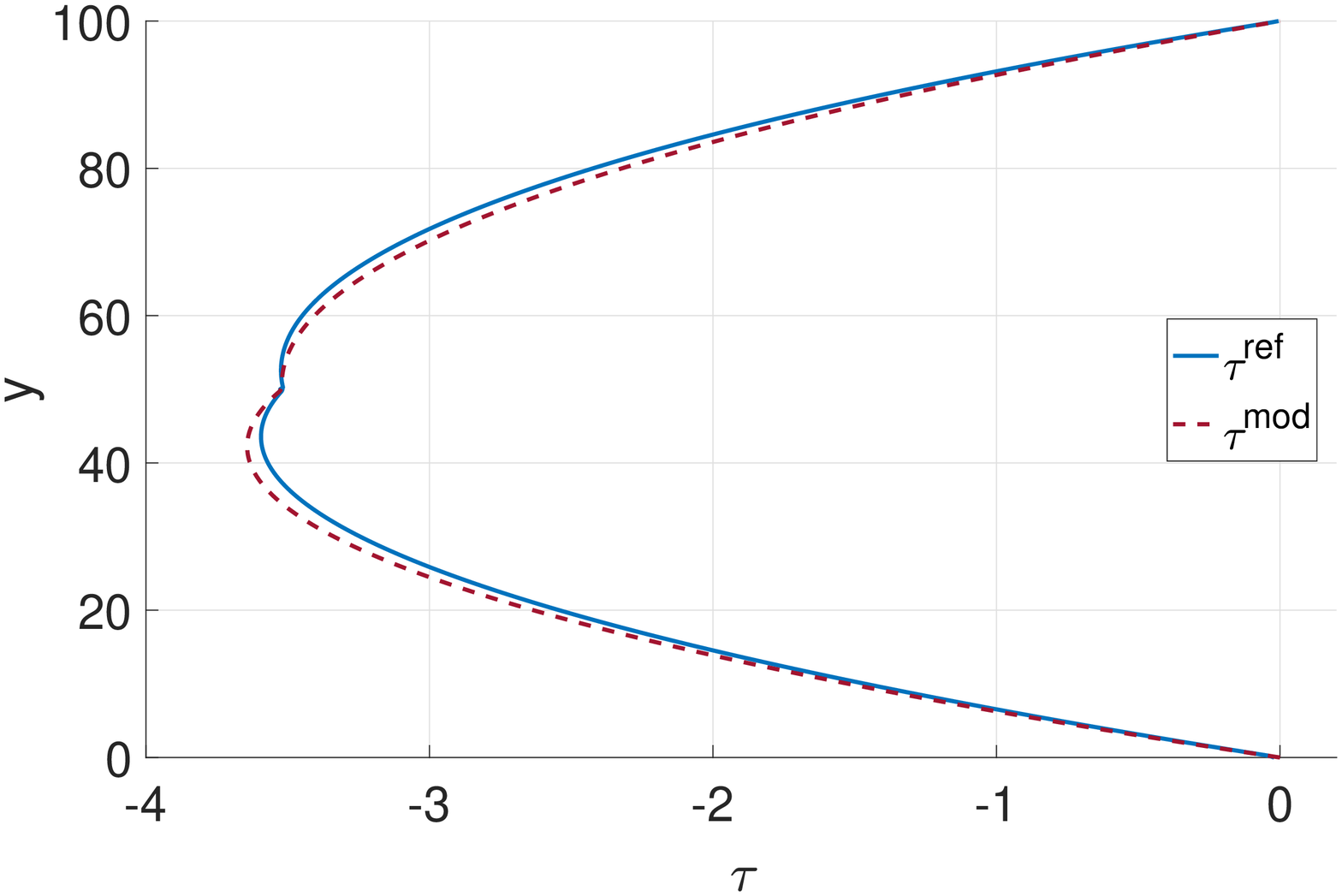}}
\subfigure[shear stress components, $x= 375$]{
\label{f_layer+_t3_par}
\includegraphics[width=0.48\textwidth]{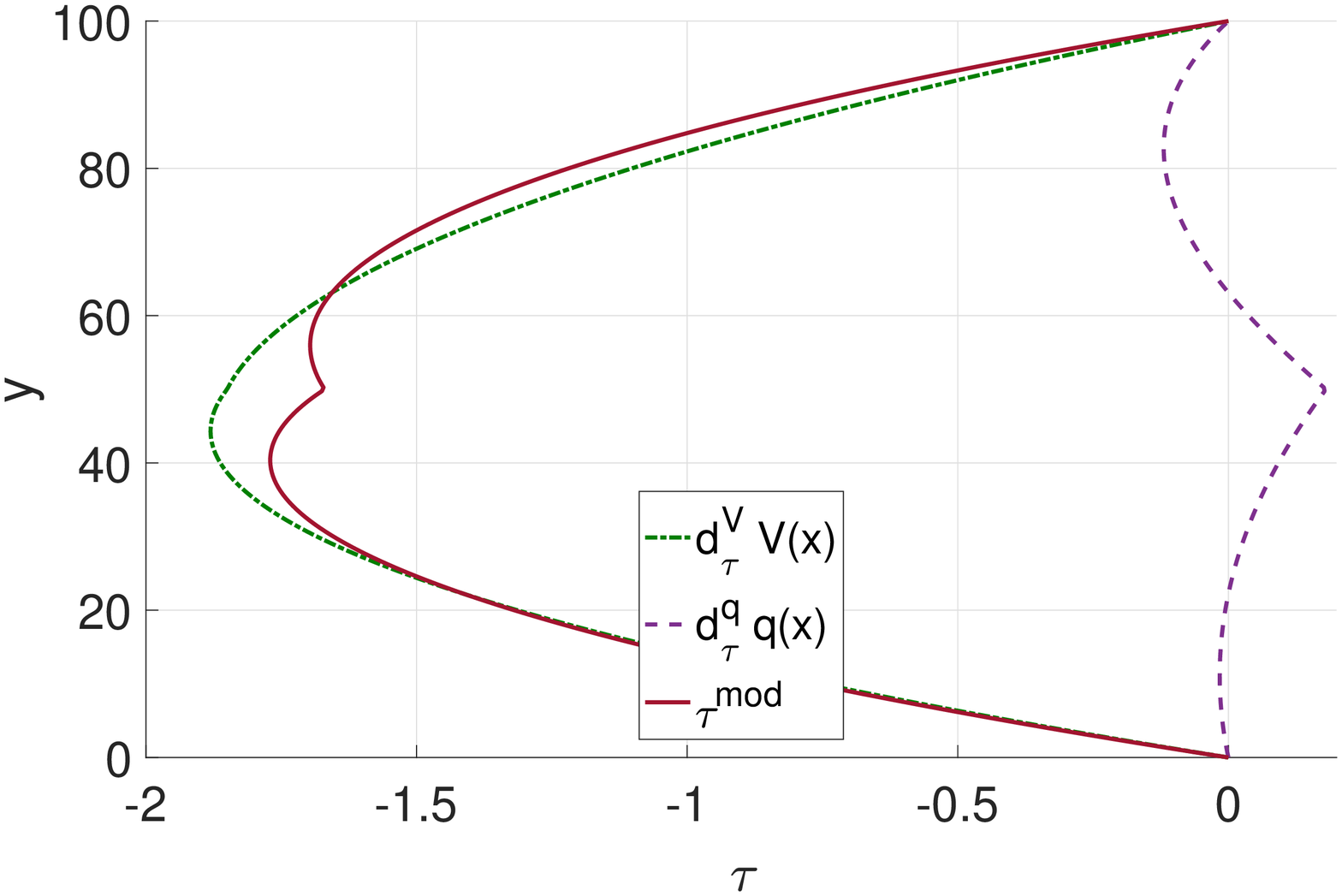}}
\subfigure[shear stress, $x= 375$]{
\label{f_layer+_t3}
\includegraphics[width=0.48\textwidth]{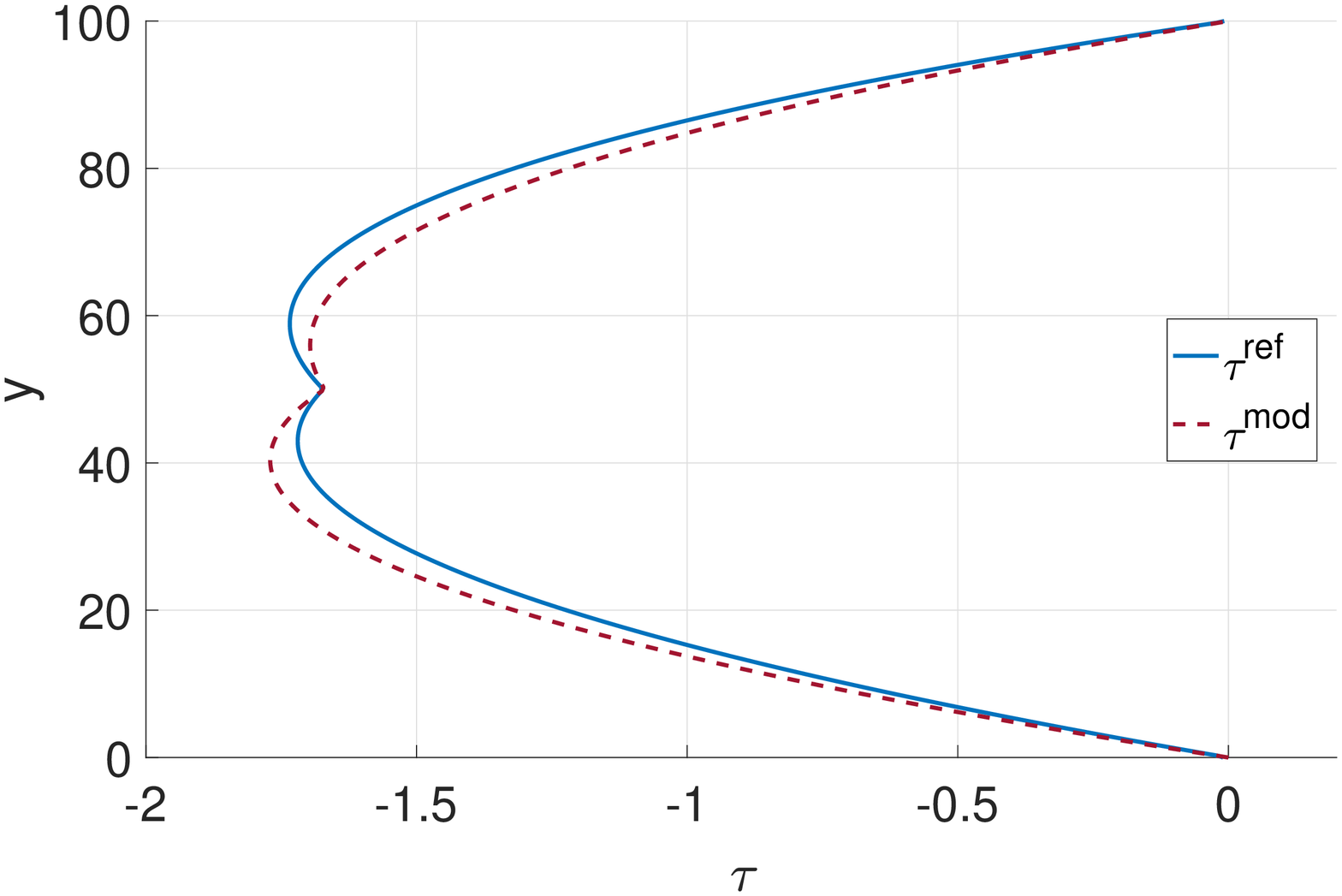}}
\caption{\footnotesize{
Bi-layer anisotropic cantilever ($\theta = 15 \, \deg$).
Shear stress distributions evaluated at $x=250 \, \milli\meter$ (Figures \ref{f_layer+_t2_par} and \ref{f_layer+_t2}), and $x=375 \, \milli\meter$ (Figures \ref{f_layer+_t3_par} and \ref{f_layer+_t3}).
Analysis of the shear stress components according to the proposed beam model (Figures \ref{f_layer+_t2_par} and \ref{f_layer+_t3_par}) and comparisons of the beam model $\psi^{mod}$ and the reference $\psi^{ref}$ solutions (Figures \ref{f_layer+_t2} and\ref{f_layer+_t3}).}} 
\label{f_layer+_tau}
\end{figure}

Figures \ref{f_layer+_sx} and \ref{f_layer+_tau} report cross-section stress distributions at $1/2l = 250 \, \milli\meter$ and $3/4l = 375 \, \milli\meter$.

Figures \ref{f_layer+_sx2_par} and \ref{f_layer+_sx3_par} highlight that the axial stress depending on the transversal internal force $d_{\sigma_x}^V \left(y\right) V \left(x\right)$ is not negligible at all, but can increase the magnitude of maximal stress up to $30 \, \%$.
Conversely, the effect of the axial stress depending on the transversal load $d_{\sigma_x}^{q} \left(y\right) q$ is less significant.
The comparison with reference solution (Figures \ref{f_layer+_sx2} and \ref{f_layer+_sx3}) reveals that the stress recovery developed in Section \ref{s_stress_rec} provides accurate estimations of the stress magnitude, with relative errors rarely bigger than $10 \, \%$.
In particular, the stress recovery correctly predicts the jump of axial stress at the interlayer surface.
Conversely, reference solution reveals the presence of some higher order effects near to the free end of the cantilever that the beam model is not able to catch (see Remark \ref{r_bondary_effects}), and may locally lead to an increase of the relative errors up to $40 \, \%$.

Figures \ref{f_layer+_t2_par} and \ref{f_layer+_t3_par} highlight that the shear stress depending on transversal load $d_{\tau}^q \left(y\right) q$ is not negligible, but it can lead to the creation of a local minimum on the interlayer surface.
The comparison with reference solution (Figures \ref{f_layer+_t2_par} and \ref{f_layer+_t3_par}) reveals that the proposed stress recovery is in good agreement with reference solution, leading to relative errors rarely bigger than $5 \, \%$.

In order to complete the discussion of the proposed model capabilities, Tables \ref{t_u_errors}, \ref{t_phi_errors}, and  \ref{t_v_errors} compare the solutions obtained using different models. 
Maximal displacements $\psi \left(l\right)$, with  $\psi = u, \phi, v$, are evaluated using 2D \ac{FE} $\psi^{ref} \left(l\right)$, standard \ac{EB} beam model $\psi^{EB}\left(l\right)$, and the proposed beam model $\psi^{mod}\left(l\right)$.
Only for the transversal displacement, also the Timoshenko beam model is considered since its solution differs form \ac{EB} ($v^{EB}\left(l\right) + v^{T}\left(l\right)$.
Relative errors are computed as 
\begin{equation}
\label{rel_error}
e^{i} = \frac{\left| \psi^{i} \left(l\right) - \psi^{ref} \left(l\right) \right|}{ \left| \psi^{ref} \left(l\right) \right|} \mbox{ with } i = EB, T, mod
\end{equation}
Finally, numerical results and relative errors are provided for $\lambda = 5, \, 10, \mbox{ and } 20$ and $\theta = \pm 15 \, \deg$.
\begin{table}[htbp]
\centering
\begin{tabular}{l|r|r|rr|rr|}
$\lambda$
&
$\theta \left[\deg\right]$
&
$u^{ref} \left[\milli\meter\right]$
&
$u^{EB} \left[\milli\meter\right]$
&
$u^{mod} \left[\milli\meter\right]$
&
$e^{EB}_u \left[\%\right]$
&
$e^{mod}_u \left[\%\right]$
\\
\hline
5 & +15 &
 1.173e--1 & 0.000e+0 & 1.078e--1 & 100 & 8.10 \\
5 & --15 &
 --1.125e--1 & 0.000e+0 & --1.078e--1 & 100 & 4.18 \\
10 & +15 &
 4.612e--1 & 0.000e+0 & 4.311e--1 & 100 & 6.53 \\
10 & --15 &
 --4.504e--1 & 0.000e+0 & --4.311e--1 & 100 & 4.29 \\
20 & +15 &
 1.828e+0 & 0.000e+0 & 1.724e+0 & 100 & 5.69 \\
20 & --15 &
 --1.803e+0 & 0.000e+0 & --1.724e+0 & 100 & 4.38
\end{tabular}
\caption{\footnotesize{
Bi-layer anisotropic cantilever. 
Maximal axial displacement $u \left(l\right)$ evaluated according to \ac{EB} $u^{EB}$ and proposed $u^{mod}$ beam models and relative errors.}} 
\label{t_u_errors}
\end{table}
\begin{table}[htbp]
\centering
\begin{tabular}{l|r|r|rr|rr|}
$\lambda$
&
$\theta \left[\deg\right]$
&
$\phi^{ref} \left[\rad\right]$
&
$\phi^{EB} \left[\rad\right]$
&
$\phi^{mod} \left[\rad\right]$
&
$e^{EB}_{\phi} \left[\%\right]$
&
$e^{mod}_{\phi} \left[\%\right]$
\\
\hline
5 & +15 &
 --3.569e--2 & --3.189e--2 & --3.513e--2 & 10.6 & 1.57 \\
5 & --15 &
 --2.841e--2 & --3.189e--2 & --2.864e--2 & 12.2 & 0.81 \\
10 & +15 &
 --2.706e--1 & --2.551e--1 & --2.681e--1 & 5.73 & 0.92 \\
10 & --15 &
 --2.410e--1 & --2.551e--1 & --2.421e--1 & 5.85 & 0.46 \\
20 & +15 &
 --2.103e+0 & --2.041e+0 & --2.093e+0 & 2.95 & 0.48 \\
20 & --15 &
 --1.984e+0 & --2.041e+0 & --1.989e+0 & 2.87 & 0.25
\end{tabular}
\caption{\footnotesize{
Bi-layer anisotropic cantilever. 
Maximal rotations $\phi\left(l\right)$  evaluated according to \ac{EB} $\phi^{EB}$ and proposed $\phi^{mod}$ beam models and relative errors.}} 
\label{t_phi_errors}
\end{table}
\begin{table}[htbp]
\centering
\begin{tabular}{l|r|r|rrr|rrr|}
$\lambda$
&
$\theta \left[\deg\right]$
&
$v^{ref} \left[\milli\meter\right]$
&
$v^{EB} \left[\milli\meter\right]$
&
$v^{EB} + v^{T} \left[\milli\meter\right]$
&
$v^{mod} \left[\milli\meter\right]$
&
$e^{EB}_v \left[\%\right]$
&
$e^{EB+T}_v \left[\%\right]$
&
$e^{mod}_v \left[\%\right]$
\\
\hline
5 & +15 &
--1.545e+1 & --1.196e+1 & --1.364e+1 & --1.516e+1 & 22.6 & 11.7 & 1.88 \\
5 & --15 &
--1.177e+1 & --1.196e+1 & --1.364e+1 & --1.191e+1 & 1.61 & 15.9 & 1.19 \\
10 & +15 &
--2.130e+2 & --1.913e+2 & --1.981e+2 & --2.106e+2 & 10.2 & 7.00 & 1.13 \\
10 & --15 &
--1.834e+2 & --1.913e+2 & --1.981e+2 & --1.847e+2 & 4.31 & 8.02 & 0.71 \\
20 & +15 &
--3.210e+3 & --3.061e+3 & --3.088e+3 & --3.190e+3 & 4.64 & 3.80 & 0.62 \\
20 & --15 &
--2.972e+3 & --3.061e+3 & --3.088e+3 & --2.982e+3 & 2.99 & 3.90 & 0.34
\end{tabular}
\caption{\footnotesize{
Bi-layer anisotropic cantilever. 
Maximal transversal displacements $v\left(l\right)$ evaluated according to \ac{EB} $v^{EB}$, Timoshenko $v^{T}$, and proposed $v^{mod}$ beam models and relative errors.}} 
\label{t_v_errors}
\end{table}

On the one hand, relative errors decrease increasing the slenderness for all the considered beam models, consistently with standard beam model assumptions.
On the other hand, it is worth highlighting that \ac{EB} and Timoshenko beam models lead to errors that are often greater than $10 \, \%$ and are therefore not acceptable for most of engineering applications.
Conversely, the proposed beam model leads to errors that are usually smaller than $5 \, \%$, reaching an accuracy adequate for most of engineering applications.

Furthermore, Table \ref{t_v_errors} highlights that Timoshenko beam model not always performs better than \ac{EB}, even considering tick beams.
In particular, for $\lambda = 5$ and $\theta = - 15 \, \deg$ maximal transversal displacement predicted by the proposed beam model qualitatively coincides with the \ac{EB} solution i.e., $v^{mod} \left(l\right) \approx v^{EB} \left(l\right)$.
On the one hand, such a result highlights (i) the deep influence of fiber direction on the structural element stiffness (see Remarck \ref{r_D_even_odd}) and (ii) the inappropriateness of beam models developed for isotropic structural element.
On the other hand, the extremely low errors obtained for both positive and negative $\theta$ confirm the effectiveness of the proposed model in handling all peculiar aspects of anisotropic beams.

\begin{figure}[htbp]
\centering 
\includegraphics[width=0.5\columnwidth]{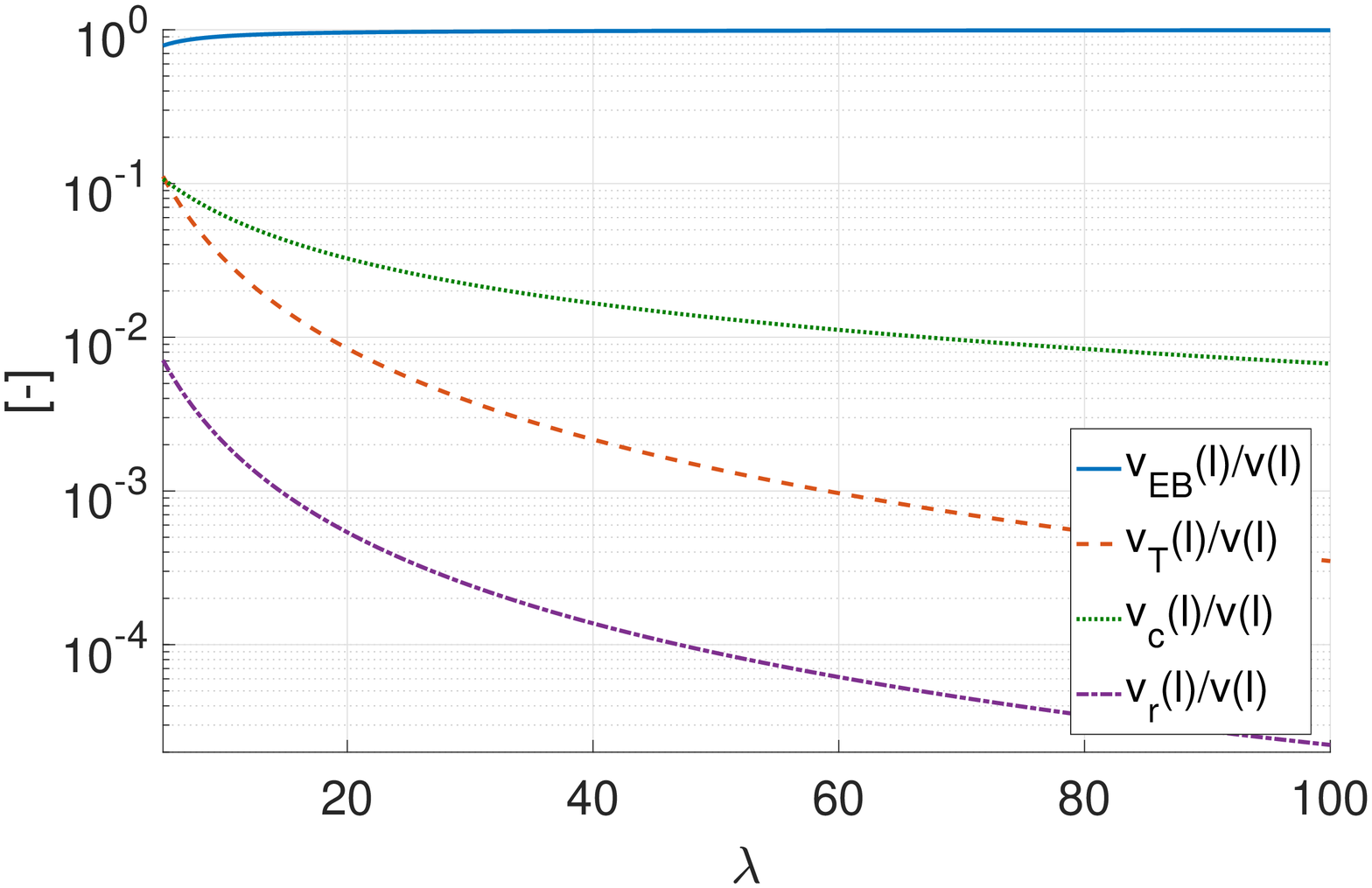}
\caption{\footnotesize{
Bi-layer anisotropic cantilever ($\theta = 15 \, \deg$). 
Incidence of the maximal transversal displacement components $v_{EB} \left(l\right)$, $v_{T} \left(l\right)$, $v_{c} \left(l\right)$, and  $v_{r} \left(l\right)$ evaluated for varying $\lambda$.}} 
\label{f_percentage}
\end{figure}

Figure \ref{f_percentage} reports the weight of the four components $v_{EB} \left(l\right)$, $v_{T} \left(l\right)$, $v_{c} \left(l\right)$, and $v_{r} \left(l\right)$ on the total transversal displacement as a function of $\lambda$.
The analysis is limited to geometry and material mechanical properties introduced at the beginning of Section \ref{s_num_res}, but it highlights several effects of the anisotropy on the structural response of beams.
The component $v_{c} \left(l\right)$, depending on the material coupling term $G_x$, is always bigger than component depending of shear $v_{T} \left(l\right)$.
As an example, considering $\lambda = 20$ $v_{c} \left(l\right) / v \left(l\right) >3 \, \%$ whereas $v_{T} \left(l\right) / v \left(l\right) <1 \%$.
Furthermore, for $\lambda = 80$ $v_{c} \left(l\right) / v\left(l\right) \approx 1 \, \%$ whereas $v_{T} \left(l\right) / v \left(l\right) \approx 0.05 \, \%$.
As a consequence, it is possible to conclude that the material coupling contribution $v_{c} \left(x\right)$ could be significant also for slender structural elements for which, instead, shear contribution is negligible.
Conversely, the transversal displacement $v_{r} \left(l\right)$ always contributes to the total displacement less than $1 \, \%$.

\subsection{Doubly-clamped beam}
\label{s_clamp_clamp}

This section considers the statically indeterminate multilayer beam depicted in Figure~\ref{f_cantilever}, aiming at confirming the capabilities of the proposed beam model in effectively estimating anisotropic beam stiffness.
Analytical solution reported in Equation \eqref{ode_sol} is still valid.
Conversely the following \acp{BC} has to be considered:
\begin{equation}
u  \left( 0 \right) = u  \left( l \right) = 0; \quad
\phi \left( 0 \right) = \phi \left( l \right) = 0; \quad
v \left( 0 \right) = v \left( l \right) = 0
\end{equation}
Analytical expression for the coefficients $C_i$ for $i = 1 \dots 6$ turns out to be extremely complex and, for brevity, they will not be reported.
Anyway, it has to be noticed that all the coefficients depends on all mechanical properties $E_{11}$, $G_{12}$ , and $G_x$.
As a consequence, the subdivision of displacements in components $\left(\cdot\right)_{EB}$, $\left(\cdot\right)_{T}$, $\left(\cdot\right)_{c}$, and $\left(\cdot\right)_{r}$ introduced in Equation \eqref{ode_sol} is no longer meaningful and it will not be considered in the following.
We set $l = 1000 \, \milli\meter$ and we use the geometrical and mechanical properties reported in Equations \eqref{mech_prop} and 
\eqref{mech_prop1}.

\begin{figure}[htbp]
\centering 

\subfigure[axial internal force]{
\label{f_clamp+_N}
\includegraphics[width=0.48\textwidth]{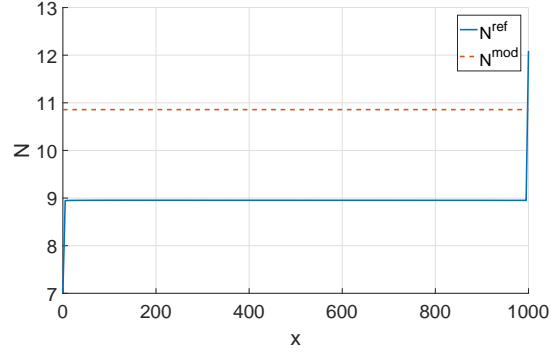}}

\subfigure[bending moment]{
\label{f_clamp+_M}
\includegraphics[width=0.48\textwidth]{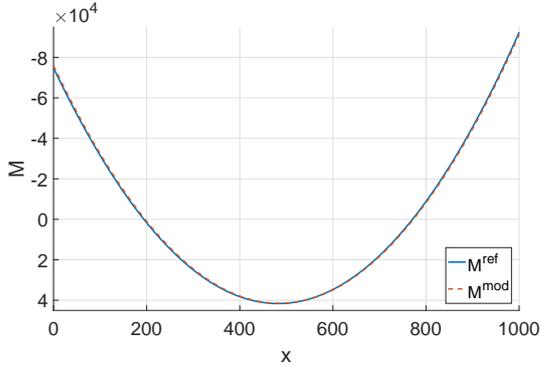}}
\subfigure[transversal internal force]{
\label{f_clamp+_V}
\includegraphics[width=0.48\textwidth]{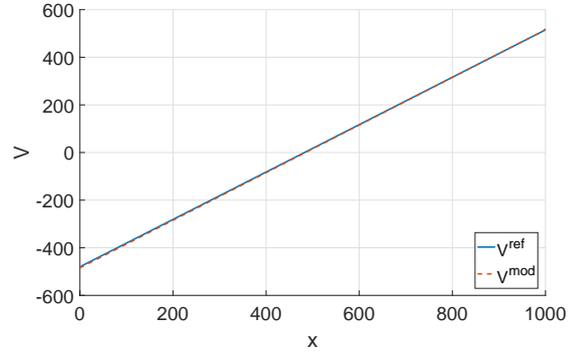}}
\caption{\footnotesize{
Doubly clamped bi-layer anisotropic beam.
Analysis of internal forces. 
Comparisons of the beam model $\psi^{mod}$ and the reference $\psi^{ref}$ solutions.}} 
\label{f_clamp+_int_for}
\end{figure}

Figure \ref{f_clamp+_int_for} reports numerical results concerning distribution of internal forces $N \left(x\right)$, $M \left(x\right)$, and $V \left(x\right)$.
Numerical results highlight a non-trivial effect of the material anisotropy. 
The distribution of internal forces is non-symmetric and reactions on the right hand side clamp are greater than the ones in the left hand side clamp despite \acp{BC} and load are symmetric with respect to the beam mid-span. 
Once more, comparison with reference solution reveals the high accuracy of the proposed beam model that predicts both bending moment and transversal internal force with negligible errors.
Finally, the proposed beam model correctly predicts a non-vanishing, constant distribution of axial internal force, which magnitude is anyway negligible if compared with transversal internal force and bending moment.

Tables \ref{t_N_errors}, \ref{t_M_errors}, and \ref{t_V_errors} report constraint reactions (i.e., $N\left(x\right)$, $M\left(x\right)$, and $V\left(x\right)$ for $x = 0, l$) evaluated using 2D \ac{FE} $\psi^{ref} \left(l\right)$, standard \ac{EB} beam model $\psi^{EB}\left(l\right)$, and the proposed beam model $\psi^{mod}\left(l\right)$ for $\lambda = 5, \, 10, \mbox{ and } 20$.
Relative errors are computed according to Equation \eqref{rel_error}. 
\begin{table}[htbp]
\centering
\begin{tabular}{l|r|r|rr|rr|}
$\lambda$
&
$x$
&
$N^{ref} \left[\newton\right]$
&
$N^{EB} \left[\newton\right]$
&
$N^{mod} \left[\newton\right]$
&
$e^{EB}_N \left[\%\right]$
&
$e^{mod}_N \left[\%\right]$
\\
\hline
5 & 0 &
7.356e+0 & 0.000e+0 & 8.742e+0 & 100 & 18.8 \\
5 & $l$ &
7.356e+0 & 0.000e+0 & 8.742e+0 & 100 & 18.8 \\
10 & 0 &
8.959e+0 & 0.000e+0 & 1.092e+1 & 100 & 21.9 \\
10 & $l$ &
8.959e+0 & 0.000e+0 & 1.092e+1 & 100 & 21.9 \\
20 & 0 &
9.472e+0  & 0.000e+0 & 1.165e+1 & 100 & 23.0 \\
20 & $l$ &
9.472e+0  & 0.000e+0 & 1.165e+1 & 100 & 23.0
\end{tabular}
\caption{\footnotesize{
Doubly clamped bi-layer anisotropic beam. 
Axial constraint reactions $N \left(0\right)$ and $N \left(l\right)$ evaluated according to \ac{EB} $N^{EB}$ and proposed $N^{mod}$ beam models and relative errors.}} 
\label{t_N_errors}
\end{table}

\begin{table}[htbp]
\centering
\begin{tabular}{l|r|r|rr|rr|}
$\lambda$
&
$x$
&
$M^{ref} \left[\newton \meter\right]$
&
$M^{EB} \left[\newton \meter\right]$
&
$M^{mod} \left[\newton \meter\right]$
&
$e^{EB}_{M} \left[\%\right]$
&
$e^{mod}_{M} \left[\%\right]$
\\
\hline
5 & 0 &
--1.732e+4  & --2.083e+4 & --1.794e+4 & 20.3 & 3.58 \\
5 & $l$ &
--2.482e+4  & --2.083e+4 & --2.415e+4 & 16.1 & 2.70 \\
10 & 0 &
--7.458e+4 & --8.333e+4 & --7.583e+4 & 11.7 & 1.68 \\
10 & $l$ &
--9.244e+4  & --8.333e+4 & --9.137e+4 & 9.86 & 1.16 \\
20 & 0 &
--3.138e+5  & --3.333e+5 & --3.170e+5 & 6.21 & 1.02 \\
20 & $l$ &
--3.526e+5  & --3.333e+5 &  --3.502e+5 & 5.47 & 0.68
\end{tabular}
\caption{\footnotesize{
Doubly clamped bi-layer anisotropic beam. 
Bending moment constraint reactions $M \left(0\right)$ and $M \left(l\right)$ evaluated according to \ac{EB} $M^{EB}$ and proposed $M^{mod}$ beam models and relative errors.}} 
\label{t_M_errors}
\end{table}

\begin{table}[htbp]
\centering
\begin{tabular}{l|r|r|rr|rr|}
$\lambda$
&
$x$
&
$V^{ref} \left[\newton\right]$
&
$V^{EB} \left[\newton\right]$
&
$V^{mod} \left[\newton\right]$
&
$e^{EB}_V \left[\%\right]$
&
$e^{mod}_V \left[\%\right]$
\\
\hline
5 & 0 &
--2.333e+2 & --2.500e+2 & --2.376e+2 & 7.16 & 1.84 \\
5 & $l$ &
--2.645e+2 & --2.500e+2 & --2.624e+2 & 5.48 & 0.79 \\
10 & 0 &
--4.817e+2 & --5.000e+2 & --4.845e+2 & 3.80 & 0.58 \\
10 & $l$ &
--5.184e+2 & --5.000e+2 & --5.155e+2 & 3.55 & 0.56 \\
20 & 0 &
--9.921e+2 & --1.000e+3 & --9.834e+2 & 0.80 & 0.88 \\
20 & $l$ &
--1.026e+2 & --1.000e+3 & --1.017e+3 & 2.53 & 0.88
\end{tabular}
\caption{\footnotesize{
Doubly clamped bi-layer anisotropic beam. 
Shear constraint reactions $V \left(0\right)$ and $V \left(l\right)$  evaluated according to \ac{EB} $V^{EB}$ and proposed $V^{mod}$ beam models and relative errors.}} 
\label{t_V_errors}
\end{table}
As already remarked in Section \ref{s_cantilever}, \ac{EB} beam model leads to errors that are often bigger than $10 \, \%$, leading to estimations that are too coarse for most of engineering applications.
Conversely, the proposed beam model leads to errors that are generally three-six times smaller and always below $5 \, \%$.
Only Table \ref{t_N_errors} highlights that model estimates axial internal force $N \left(x\right)$ with errors over $20 \, \%$.
Anyway, since the magnitude of axial internal force is approximatively 50 times smaller than transversal one, the relative error on axial internal force might not have a deep influence on the global response of structural element.

\section{Conclusions} 
\label{s_conclusion}

This paper has proposed a simple beam model that effectively handles the influence of anisotropy on the beam constitutive relations and the stress distribution.
The independent variables of the model are the internal forces and the standard Timoshenko kinematic parameters.
Despite its simplicity, the beam model has allowed to highlight the following peculiarities of anisotropic beams.
\begin{enumerate}
\item 
Material anisotropy leads transversal internal force to contribute up to $30 \, \%$ of the magnitude of axial stress, deeply affecting also the beam strength, not explicitly considered in this paper.
\item 
In beam constitutive relations, non-vanishing out-of-diagonal terms that relate transversal internal force with curvature (and bending moment with shear strain) exist and deeply influence the response of the structural element.
\item 
In addition to the standard bending contribution (proportional to cube beam-slenderness) and the shear one (proportional to beam-slenderness), a third term, depending on material coupling term and proportional to square beam-slenderness, contributes to transversal displacement.
\item 
The contribution depending on material coupling terms can be bigger than the contribution given by shear deformation and it may be non-negligible for length vs thickness ratios greater than fifty.
\end{enumerate}

A systematic comparison with analytical results and 2D \ac{FE} solutions, obtained using highly refined meshes, demonstrates the effectiveness of the proposed modeling approach.
In general, the proposed beam model has a computational cost similar to simplest beam models used in engineering practice and it estimates significant displacements and internal forces with relative errors usually smaller than $5 \, \%$. 
Conversely, coarse adaptations of beam models developed for isotropic structural elements may lead to errors greater than $20 \, \%$ in the prediction of both internal forces and displacements.
Furthermore, analysis of stress distributions demonstrates that stress recovery tools developed for isotropic structural elements are no longer effective for anisotropic ones, but ad-hoc routines has to be developed.
The main limitations of the proposed model are the assumptions on kinematics that do not allow to describe higher order effects like cross-section warping and distortion as well as phenomena that occur in the neighborhood of constraints and concentrated loads.

Future research will include the application of the proposed modeling strategy to higher order planar beams and its generalization to 3D beams and plates.

\section{Acknowledgments}

This work was funded by the Austrian Science Found (FWF) [M 2009-N32].
F. Auricchio and S. Morganti would like to acknowledge the strategic theme of the University of Pavia "Virtual Modeling and Additive Manufacturing for Advanced Materials (3D@UniPV)"

\appendix

\section{ Mechanical properties coefficients }
\label{a_mat_coeff}

\begin{equation}
\mu = E_{11} \cos^{2} \left( \theta
 \right)  \left({\frac { \left( 2 \nu+1 \right) \cos^{2} \left( \theta
\right)  - 2  \nu}{E_{11}}}+
{\frac { 1}{E_{22}}} \left( \cos^{2} \left( \theta  \right) -
2  +\frac{1}{\cos^{2} \left( \theta \right) }\right)\right)+
{\frac {1 - \cos^{2} \left( \theta
\right)  }{G_{12}}}
\end{equation}

\begin{equation}
\kappa = 4 G_{12} \cos^{2} \left( \theta \right)  \left({\frac { \left( 2\nu+1 \right) \left(1 -   \cos ^{2} \left( \theta
 \right) \right)}{E_{11}}}+
{\frac {1- \left( \cos^{2}
 \left( \theta \right)  \right)  }{E_{22}}}+
{\frac {1}{G_{12}} \left( \cos^{2}
 \left( \theta \right) - 1+\frac{1}{4 \cos^{2} \left( \theta \right)   }\right)}
\right)
\end{equation}

\begin{equation}
\frac{1}{G_x} = 2 \sin \left( \theta \right) \cos \left( \theta \right) 
\left({\frac { \left(  \left( 2 \nu+1
 \right)  \cos^{2} \left( \theta \right) -\nu
 \right) }{E_{11}}}+
2 {\frac {  \left( \cos^{2} \left( \theta
 \right) -1 \right) }{E_{22}}}+
{\frac { \left(1 -2  \cos^{2}
 \left( \theta \right)  \right)  }{G_{12}}}\right)
\end{equation}

\section{ Dimensionless coefficients $P_i$ (for $i = 1 \dots 6$)}
\label{a_p}

\begin{equation}
P_1 = 
\frac{\mu}{1+ \left( \mu - 1 \right) \alpha}
\end{equation}
%
 
\begin{equation}
P_2 =
\frac{\mu \left( 1 + \left( \mu - 1 \right) \alpha \right) }{ 1 + \left( \mu - 1 \right)  \alpha \left( \left( \mu - 1 \right) \alpha^3 +
 4 \alpha^2 - 6 \alpha + 4 \right) }
\end{equation}
 
\begin{equation}
P_3 =
\frac { \left( -1+ \left( \mu-1 \right) {\alpha}^{2}+ \left( 
-4 \mu+2 \right) \alpha \right)  \left( \alpha-1 \right) ^{2
}}{\left( 1+ \left( \mu-1 \right) \alpha \right) 
  \left( 1 + \left( \mu - 1 \right)  \alpha \left( \left( \mu - 1 \right) \alpha^3 +
 4 \alpha^2 - 6 \alpha + 4 \right) \right) }
\end{equation}
 
\begin{equation}
P_4 =  \frac {{\mu}^{2}{\alpha}^{2} \left( \alpha-1 \right) ^{2}}{ \left( 1 + \left( \mu - 1 \right)  \alpha \left( \left( \mu - 1 \right) \alpha^3 +
 4 \alpha^2 - 6 \alpha + 4 \right) \right) ^{2}}
\end{equation}
 
\begin{equation}
\begin{split}
P_5 = &
\frac{1}{\left( 1 + \left( \mu - 1 \right)  \alpha \left( \left( \mu - 1 \right) \alpha^3 +
 4 \alpha^2 - 6 \alpha + 4 \right) \right) ^{2}} 
\\
& \left( \left( \mu-1 \right) ^{
2} \left( {\mu}^{2}-\kappa \right) {\alpha}^{7}- \left(3  {
\mu}^{2}- 10 \mu \kappa+ 7 \kappa \right)  \left( \mu - 1 \right) {\alpha}^{6}
\right. \\
&  + \left( 5 {\mu}^{3}+ \left( 1 -
40 \kappa \right) {\mu}^{2}+55 \mu \kappa-21\kappa \right) {\alpha}^{5}+ \left(  \left( 70 \kappa-15
 \right) {\mu}^{2}-90 \mu \kappa+35 \kappa \right) 
{\alpha}^{4}
\\
& \left. + \left(  \left( 10 - 55 \kappa \right) {\mu}^{2
}+80 \mu \kappa-35 \kappa \right) {\alpha}^{3}+\kappa  \left( 16 \mu-21 \right)  \left( \mu-1
 \right) {\alpha}^{2}+7 \kappa  \left( \mu-1 \right) \alpha+ \kappa \right)
\end{split}
\end{equation}

\begin{equation}
\begin{split}
P_6 = &
\frac{\left( \alpha-1 \right) ^{3}}{ \left(\left( \mu-1 \right) \alpha \left( \left( \mu-1 \right) {\alpha}^{3}+ 4 {\alpha}^{2}-6  {\alpha}+ 4 \right) + 1 \right) ^{3}\mu } \left(1+ \left( 6 \mu-1 \right)  \left( \mu-1 \right) ^{3}{\alpha}^{8} \right.
\\
&\left. + \left( -42 {\mu}^{4}+129 {\mu}^{3}-140 {\mu}^{2}+61 \mu-8 \right) {\alpha}^{7}  + \left( 96 {\mu}^{4}-372 {\mu}^{3}+429 {\mu}^{2}-181 \mu+28 \right) {\alpha}^{6}\right.
\\
& \left.+ \left( 622 {\mu}^{3}-736 {\mu}^{2}+305 \mu-56 \right) {\alpha}^{5} + \left( -609 {\mu}^{3}+719 {\mu}^{2}-315 \mu+70 \right) {\alpha}^{4} \right.
\\
& \left.+ \left( 249 {\mu}^{3}-372 {\mu}^{2}+199 \mu-56 \right) {\alpha}^{3}  + \left( 79 {\mu}^{2}-71 \mu+28 \right) {\alpha}^{2}+
 \left( 11 \mu-8 \right) \alpha
\right)
\end{split}
\end{equation}

\section{Maximal displacement coefficients $Q_i$ (for $i = 1 \dots 3$)}
\label{a_q_c}

\begin{equation}
\begin{split}
Q_1 = & \frac { \mu }{ \left(  1 + \left( \mu - 1 \right)  \alpha \left( \left( \mu - 1 \right) \alpha^3 +
 4 \alpha^2 - 6 \alpha + 4 \right)   \right) ^{2}}  
\left( 
\left( \mu - 1 \right)^3 { \alpha}^{5}+ 5 \left( \mu - 1 \right)^2 { \alpha}^{4} 
\right.
\\
& \left.
+ 2  \left( -3  {\mu}^{2}+8 \mu-5  \right) { \alpha}^{3}+ 2 \left( 2 \mu - 5 \right)  \left( \mu-1 \right) { \alpha}^{2}+ 5 \left( \mu-1 \right)  \alpha + 1 \right)
\end{split}
\end{equation}
 
\begin{equation}
\begin{split}
Q_2 = & \frac {   1}{ 5\left(   1 + \left( \mu - 1 \right)  \alpha \left( \left( \mu - 1 \right) \alpha^3 +
 4 \alpha^2 - 6 \alpha + 4 \right)   \right) ^{2}} 
\left( \left( \mu-1 \right) ^{2} \left( {\mu}^{2}- \kappa \right)   { \alpha}^{7}-
\left( 3 {\mu}^{2}-20 \mu  \kappa+7  \kappa \right)  \left( \mu-1 \right)   { \alpha}^{6}
\right.
\\
& \left.
+ \left( 5 {\mu}^{3} -   \left( 40 \kappa - 1 \right)   {\mu}^{2}+55 \mu  \kappa  -21  \kappa \right) { \alpha}^{5}+ 
5 \left( \left( 14 \kappa-3 \right) {\mu}^{2}-18 \mu  \kappa  +7 \kappa   \right) { \alpha}^{4}-
\right.
\\
& \left.
5 \left( \left( 11 \kappa - 2 \right)   {\mu}^{2} - 16 \mu  \kappa + 7 \kappa \right) { \alpha}^{3}+
 \kappa \left( \mu-1 \right)^2 { \alpha}^{2}+
7  \kappa    \left( \mu-1 \right)  \alpha+ \kappa   \right) 
\end{split}
\end{equation}
 
\begin{equation}
Q_3 = \frac { -\mu^2 \alpha^2  \left( \alpha- 1 \right)^2 }{ \left(  1 + \left( \mu - 1 \right)  \alpha \left( \left( \mu - 1 \right) \alpha^3 +
 4 \alpha^2 - 6 \alpha + 4 \right)   \right) ^{2}}
\end{equation}

\begin{equation}
\begin{split}
Q_4 = & 
\frac { \left( \alpha-1 \right) ^{3} }{ 10 \left(\left( \mu-1 \right) \alpha \left( \left( \mu-1 \right) {\alpha}^{3}+ 4 {\alpha}^{2}-6  {\alpha}+ 4 \right) + 1 \right) ^{3}\mu}
\left( 1+  \left(6 \mu-1\right)  \left( \mu-1 \right) ^{3}{\alpha}^{8} 
\right.
\\
& \left. 
+ \left( -42 {\mu}^{4}+129 {\mu}^{3}-140 {\mu}^{2}+61 \mu-8 \right) {\alpha}^{7}+ \left( 96 {\mu}^{4}-372 {\mu}^{3}+429 {\mu}^{2}-181 \mu+28 \right) {\alpha}^{6}
\right.
\\
& \left. 
+ \left( 622 {\mu}^{3}-736 {\mu}^{2}+305 \mu-56 \right) {\alpha}^{5}+ \left( -609 {\mu}^{3}+719 {\mu}^{2}-315 \mu+70 \right) {\alpha}^{4}
\right.
\\
& \left. 
+ \left( 249 {\mu}^{3}-372 {\mu}^{2}+199 \mu-56 \right) {\alpha}^{3}+ \left( 79 {\mu}^{2}-71 \mu+28 \right) {{\it 
aa}}^{2}+ \left( 11 \mu-8 \right) \alpha\right)
\end{split}
\end{equation}


\begin{thebibliography}{}

\bibitem[\protect\citeauthoryear{Balduzzi, Aminbaghai, Auricchio, and
  F{\"u}ssl}{Balduzzi et~al.}{2018}]{baaf_17}
Balduzzi, G., M.~Aminbaghai, F.~Auricchio, and J.~F{\"u}ssl (2018).
\newblock Planar {Timoshenko}-like model for multilayer non-prismatic beams.
\newblock {\em International Journal of Mechanics and Materials in
  Design\/}~{\em 14\/}(1), 51--70.

\bibitem[\protect\citeauthoryear{Balduzzi, Aminbaghai, and F{\"u}ssl}{Balduzzi
  et~al.}{2017}]{baf_17}
Balduzzi, G., M.~Aminbaghai, and J.~F{\"u}ssl (2017).
\newblock Linear response of a planar {FGM} beam with non-linear variation of
  the mechanical properties.
\newblock In A.~G{\"u}emes, A.~Benjeddou, J.~Rodellar, and J.~Leng (Eds.), {\em
  SMART 2017}, pp.\  1285--1294. CIMNE.

\bibitem[\protect\citeauthoryear{Balduzzi, Aminbaghai, Sacco, F{\"u}ssl,
  Eberhardsteiner, and Auricchio}{Balduzzi et~al.}{2016}]{basfea_16}
Balduzzi, G., M.~Aminbaghai, E.~Sacco, J.~F{\"u}ssl, J.~Eberhardsteiner, and
  F.~Auricchio (2016).
\newblock Non-prismatic beams: a simple and effective {Timoshenko}-like model.
\newblock {\em International Journal of Solids and Structures\/}~{\em 90},
  236--250.

\bibitem[\protect\citeauthoryear{Balduzzi, Kandler, and F\"{u}ssl}{Balduzzi
  et~al.}{2018}]{bkf_18}
Balduzzi, G., G.~Kandler, and J.~F\"{u}ssl (2018).
\newblock Estimation of {GLT} beam stiffness based on homogenized board
  mechanical properties and composite beam theory.
\newblock In {\em Proceedings of the 6th European Conference on Computational
  Mechanics (ECCM 6)}, Glasgow, UK.

\bibitem[\protect\citeauthoryear{Balduzzi, Sacco, Auricchio, and
  F{\"u}ssl}{Balduzzi et~al.}{2017}]{bsaf_17}
Balduzzi, G., E.~Sacco, F.~Auricchio, and J.~F{\"u}ssl (2017).
\newblock Non-prismatic thin-walled beams: critical issues and effective
  modeling.
\newblock In L.~Ascione, V.~Berardi, L.~Feo, F.~Fraternali, and A.~M. Tralli
  (Eds.), {\em AIMETA2017 {XXIII} conference of the Italian Association of
  Theoretical and Applied ~Mechanics}.

\bibitem[\protect\citeauthoryear{Bauchau}{Bauchau}{1985}]{b_85}
Bauchau, O.~A. (1985).
\newblock A beam theory for anisotropic materials.
\newblock {\em Journal of Applied Mechanics\/}~{\em 52\/}(2), 416--422.

\bibitem[\protect\citeauthoryear{Bruhns}{Bruhns}{2003}]{b_03}
Bruhns, O.~T. (2003).
\newblock {\em Advanced Mechanics of Solids}.
\newblock Springer.

\bibitem[\protect\citeauthoryear{Carrera and Ciuffreda}{Carrera and
  Ciuffreda}{2005}]{cc_05}
Carrera, E. and A.~Ciuffreda (2005).
\newblock A unified formulation to assess theories of multilayered plates for
  various bending problems.
\newblock {\em Composite structures\/}~{\em 69}, 271--293.

\bibitem[\protect\citeauthoryear{Choi and Horgan}{Choi and
  Horgan}{1977}]{ch_77}
Choi, I. and C.~Horgan (1977).
\newblock Saint-venant’s principle and end effects in anisotropic elasticity.
\newblock {\em Journal of Applied Mechanics\/}~{\em 44\/}(3), 424--430.

\bibitem[\protect\citeauthoryear{Dassault Systemes, 2014}{Dassault Systemes,
  2014}{2014}]{abaqus}
Dassault Systemes, 2014 (2014).
\newblock {\em Abaqus{/CAE} User's Guide - Release 6.16}.
\newblock Providence, RI, USA.: Dassault Systemes, 2014.

\bibitem[\protect\citeauthoryear{Dong, Alpdogan, and Taciroglu}{Dong
  et~al.}{2010}]{dat_10}
Dong, S., C.~Alpdogan, and E.~Taciroglu (2010).
\newblock Much ado about shear correction factors in {Timoshenko} beam theory.
\newblock {\em International Journal of Solids and Structures\/}~{\em 47},
  1651--1665.

\bibitem[\protect\citeauthoryear{Dong, Kosmatka, and Lin}{Dong
  et~al.}{2001}]{dkl_01a}
Dong, S.~B., J.~B. Kosmatka, and H.~C. Lin (2001).
\newblock On {Saint-Venant}'s problem for an inhomogeneous, anisotropic
  cylinder - part {I}: methodology for {Saint-Venant} solutions.
\newblock {\em ASME, Journal of Applied Mechanics\/}~{\em 68}, 376--381.

\bibitem[\protect\citeauthoryear{Dufour, Antolin, Sangalli, Auricchio, and
  Reali}{Dufour et~al.}{2018}]{dasar_17}
Dufour, J.-E., P.~Antolin, G.~Sangalli, F.~Auricchio, and A.~Reali (2018).
\newblock A cost-effective isogeometric approach for composite plates based on
  a stress recovery procedure.
\newblock {\em Composites Part B: Engineering\/}~{\em 138}, 12--18.

\bibitem[\protect\citeauthoryear{Groh and Weaver}{Groh and
  Weaver}{2014}]{gw_14}
Groh, R. and P.~Weaver (2014).
\newblock Buckling analysis of variable angle tow, variable thickness panels
  with transverse shear effects.
\newblock {\em Composite Structures\/}~{\em 107}, 482--493.

\bibitem[\protect\citeauthoryear{Groh and Weaver}{Groh and
  Weaver}{2016}]{gw_16a}
Groh, R. and P.~M. Weaver (2016).
\newblock A computationally efficient {2D} model for inherently equilibrated
  {3D} stress predictions in heterogeneous laminated plates. {Part I}: model
  formulation.
\newblock {\em Composite structures\/}~{\em 156}, 186--217.

\bibitem[\protect\citeauthoryear{Gupta, Sarojini, Shah, and Hodges}{Gupta
  et~al.}{2018}]{gssh_18}
Gupta, M., D.~Sarojini, A.~Shah, and D.~H. Hodges (2018).
\newblock Dimensional reduction technique for analysis of aperiodic
  inhomogeneous structures.
\newblock In {\em 2018 {AIAA/ASCE/AHS/ASC} Structures, Structural Dynamics, and
  Materials Conference}.

\bibitem[\protect\citeauthoryear{Hashin}{Hashin}{1967}]{h_67}
Hashin, Z. (1967).
\newblock Plane anisotropic beams.
\newblock {\em Journal of Applied Mechanics\/}~{\em 34\/}(2), 257--262.

\bibitem[\protect\citeauthoryear{Horgan and Carlsson}{Horgan and
  Carlsson}{2018}]{hc_18}
Horgan, C.~O. and L.~A. Carlsson (2018).
\newblock {\em Comprehensive composite materials {II}}, Volume~7, Chapter
  {Saint-Venant} end effects for anisotropic materials, pp.\  38--55.
\newblock Oxfor Academic Press.

\bibitem[\protect\citeauthoryear{Jourawski}{Jourawski}{1856}]{j_56}
Jourawski, D. (1856).
\newblock Sur le r{\'e}sistance d{'}un corps prismatique et d{'}une piece
  compos{\'e}e en bois ou on t{\^o}le de fer {\`a} une force perpendiculaire
  {\`a} leur longeur.
\newblock In {\em Annales des Ponts et Chauss{\'e}es}, Volume~12, pp.\
  328--351.

\bibitem[\protect\citeauthoryear{Jung, Nagaraj, and Chopra}{Jung
  et~al.}{2002}]{jnc_02}
Jung, S.~N., V.~Nagaraj, and I.~Chopra (2002).
\newblock Refined structural model for thin- and thick- walled composite rotor
  blades.
\newblock {\em AIAA journal\/}~{\em 40\/}(1), 105--116.

\bibitem[\protect\citeauthoryear{Kandler, F{\"u}ssl, Serrano, and
  Eberhardsteiner}{Kandler et~al.}{2015}]{kfse_15}
Kandler, G., J.~F{\"u}ssl, E.~Serrano, and J.~Eberhardsteiner (2015).
\newblock Effective stiffness prediction of glt beams based on stiffness
  distributions of individual lamellas.
\newblock {\em Wood Science and Technology\/}~{\em 49\/}(6), 1101--1121.

\bibitem[\protect\citeauthoryear{Karttunen and Von~Hertzen}{Karttunen and
  Von~Hertzen}{2016}]{kh_16}
Karttunen, A.~T. and R.~Von~Hertzen (2016).
\newblock On the foundations of anisotropic interior beam theories.
\newblock {\em Composites Part B: Engineering\/}~{\em 87}, 299--310.

\bibitem[\protect\citeauthoryear{Kosmatka, Lin, and Dong}{Kosmatka
  et~al.}{2001}]{dkl_01b}
Kosmatka, J.~B., H.~C. Lin, and S.~B. Dong (2001).
\newblock On {Saint-Venant}'s problem for an inhomogeneous, anisotropic
  cylinder - part {II}: cross-sectional properties.
\newblock {\em ASME, Journal of Applied Mechanics\/}~{\em 68}, 382--391.

\bibitem[\protect\citeauthoryear{Lekhnitski{\u\i}}{Lekhnitski{\u\i}}{1968}]{l_68}
Lekhnitski{\u\i}, S. (1968).
\newblock {\em Anisotropic plates}.
\newblock Gordon and Breach.

\bibitem[\protect\citeauthoryear{Lin, Dong, and Kosmatka}{Lin
  et~al.}{2001}]{dkl_01c}
Lin, H.~C., S.~B. Dong, and J.~B. Kosmatka (2001).
\newblock On {Saint-Venant}'s problem for an inhomogeneous, anisotropic
  cylinder - part {III}: end effects.
\newblock {\em ASME, Journal of Applied Mechanics\/}~{\em 68}, 392--398.

\bibitem[\protect\citeauthoryear{Mascia and Vanalli}{Mascia and
  Vanalli}{2012}]{mv_12}
Mascia, N.~T. and L.~Vanalli (2012).
\newblock Evaluation of the coefficients of mutual influence of wood through
  off-axis compression tests.
\newblock {\em Construction and Building Materials\/}~{\em 30}, 522--528.

\bibitem[\protect\citeauthoryear{Mascia, Vanalli, Paccola, and Scoaris}{Mascia
  et~al.}{2010}]{mvps_10}
Mascia, N.~T., L.~Vanalli, R.~R. Paccola, and M.~R. Scoaris (2010).
\newblock Mechanical behaviour of wood beams with grain orientation.
\newblock {\em Mec{á}nica Computacional\/}~{\em {XXIX}}, 2839--2854.

\bibitem[\protect\citeauthoryear{Murakami, Reissner, and Yamakawa}{Murakami
  et~al.}{1996}]{mry_96}
Murakami, H., E.~Reissner, and J.~Yamakawa (1996).
\newblock Anisotropic beam theories with shear deformation.
\newblock {\em Journal of Applied Mechanics\/}~{\em 63\/}(3), 660--668.

\bibitem[\protect\citeauthoryear{Murakami and Yamakawa}{Murakami and
  Yamakawa}{1996}]{my_96}
Murakami, H. and J.~Yamakawa (1996).
\newblock On approximate solutions for the deformation of plane anisotropic
  beams.
\newblock {\em Composites Part B: Engineering\/}~{\em 27\/}(5), 493--504.

\bibitem[\protect\citeauthoryear{Pech, Kandler, Lukacevic, and F{\"u}ssl}{Pech
  et~al.}{2019}]{pklf_18}
Pech, S., G.~Kandler, M.~Lukacevic, and J.~F{\"u}ssl (2019).
\newblock Metamodel assisted optimization of glued laminated timber systems by
  reordering wooden lamellas using metaheuristic algorithms.
\newblock {\em Engineering Applications of Artificial Intelligence\/}~{\em 79},
  129--141.

\bibitem[\protect\citeauthoryear{Qin and Librescu}{Qin and
  Librescu}{2002}]{ql_02}
Qin, Z. and L.~Librescu (2002).
\newblock On a shear-deformable theory of anisotropic thin-walled beams:
  further contribution and validations.
\newblock {\em Composite Structures\/}~{\em 56\/}(4), 345--358.

\bibitem[\protect\citeauthoryear{Rajagopal}{Rajagopal}{2014}]{phd_g_14}
Rajagopal, A. (2014).
\newblock {\em Advancements in rotor blade cross-sectional analysis using the
  variational-asymptotic method}.
\newblock Ph.\ D. thesis, Georgia Institute of Technology.

\bibitem[\protect\citeauthoryear{Silvestre and Camotim}{Silvestre and
  Camotim}{2002}]{sc_02}
Silvestre, N. and D.~Camotim (2002).
\newblock First-order generalised beam theory for arbitrary orthotropic
  materials.
\newblock {\em Thin-Walled Structures\/}~{\em 40\/}(9), 755--789.

\bibitem[\protect\citeauthoryear{Tornabene, Fantuzzi, Bacciocchi, and
  Reddy}{Tornabene et~al.}{2017}]{tfbr_17}
Tornabene, F., N.~Fantuzzi, M.~Bacciocchi, and J.~Reddy (2017).
\newblock A posteriori stress and strain recovery procedure for the static
  analysis of laminated shells resting on nonlinear elastic foundation.
\newblock {\em Composites Part B: Engineering\/}~{\em 126}, 162--191.

\bibitem[\protect\citeauthoryear{Vanalli, Mascia, and Paccola}{Vanalli
  et~al.}{2003}]{vmp_03}
Vanalli, L., N.~T. Mascia, and R.~R. Paccola (2003).
\newblock Influence of the anisotropy on the mechanical behavior of laminated
  beams.
\newblock In {\em Proceedings of {COBEM 2003}}.

\bibitem[\protect\citeauthoryear{Vidal, Gallimard, and Polit}{Vidal
  et~al.}{2012}]{vgp_12}
Vidal, P., L.~Gallimard, and O.~Polit (2012).
\newblock Composite beam finite element based on the proper generalized
  decomposition.
\newblock {\em Computers \& Structures\/}~{\em 102}, 76--86.

\bibitem[\protect\citeauthoryear{Yu, Hodges, Volovoi, and Cesnik}{Yu
  et~al.}{2002}]{yhvc_02}
Yu, W., D.~H. Hodges, V.~Volovoi, and C.~E. Cesnik (2002).
\newblock On {Timoshenko}-like modeling of initially curved and twisted
  composite beams.
\newblock {\em International Journal of Solids and Structures\/}~{\em 39},
  5101--5121.

\bibitem[\protect\citeauthoryear{Yun, Volovoi, Hodges, and Hong}{Yun
  et~al.}{2002}]{yvhh_02}
Yun, W., V.~Volovoi, D.~H. Hodges, and X.~Hong (2002).
\newblock Validation of the variational asymptotic beam sectional analysis
  ({VABS}).
\newblock {\em {AIAA} (American Institute of Aeronautics and Astronautics)
  journal\/}~{\em 40}, 2105--2113.

\end{thebibliography}

\end{document}